\documentclass{aastex631}
\usepackage{multirow}
\usepackage{color}

\received{March 26, 2022}
\revised{July 10, 2022}
\accepted{Octorber 15, 2022}

\submitjournal{ApJ}

\shorttitle{Properties of AGN-obscuring silicate with full-range Spitzer/IRS spectra}
\shortauthors{T. Tsuchikawa et al. }
\graphicspath{{./}}
\begin{document}

\title{Spitzer/IRS full spectral modeling to characterize mineralogical properties of silicate dust in heavily obscured AGNs\footnote{Released on **, **, 2022}}

\correspondingauthor{H. Kaneda}
\email{kaneda@u.phys.nagoya-u.ac.jp}

\author[0000-0002-0715-8244]{T. Tsuchikawa}
\affiliation{Graduate School of Science, Nagoya University, Furo-cho, Chikusa-ku, Nagoya, Aichi 464-8602, Japan}

\author{H. Kaneda}
\affiliation{Graduate School of Science, Nagoya University, Furo-cho, Chikusa-ku, Nagoya, Aichi 464-8602, Japan}

\author{S. Oyabu}
\affiliation{Institute of Liberal Arts and Sciences, Tokushima University, 1-1 Minami-Jyosanjima, Tokushima-shi, Tokushima, 770-8502, Japan}

\author{T. Kokusho}
\affiliation{Graduate School of Science, Nagoya University, Furo-cho, Chikusa-ku, Nagoya, Aichi 464-8602, Japan}

\author{H. Kobayashi}
\affiliation{Graduate School of Science, Nagoya University, Furo-cho, Chikusa-ku, Nagoya, Aichi 464-8602, Japan}

%\author{M. Yamagishi}
%\affiliation{Institute of Astronomy, The University of Tokyo, 2-21-1 Osawa, Mitaka, Tokyo 181-0015, Japan}
%\affiliation{National Astronomical Observatory of Japan, National Institutes of Natural Sciences, 2-21-1 Osawa, Mitaka, Tokyo 181-8588, Japan}

\author{Y. Toba}
\affiliation{National Astronomical Observatory of Japan, 2-21-1 Osawa, Mitaka, Tokyo 181-8588, Japan}
\affiliation{Department of Astronomy, Kyoto University, Kitashirakawa-Oiwake-cho, Sakyo-ku, Kyoto 606-8502, Japan}
\affiliation{Academia Sinica Institute of Astronomy and Astrophysics, 11F of Astronomy-Mathematics Building, AS/NTU, No.1, Section 4, Roosevelt Road, Taipei 10617, Taiwan}
\affiliation{Research Center for Space and Cosmic Evolution, Ehime University, 2-5 Bunkyo-cho, Matsuyama, Ehime 790-8577, Japan}

%% Note that the \and command from previous versions of AASTeX is now
%% depreciated in this version as it is no longer necessary. AASTeX 
%% automatically takes care of all commas and "and"s between authors names.

%% AASTeX 6.31 has the new \collaboration and \nocollaboration commands to
%% provide the collaboration status of a group of authors. These commands 
%% can be used either before or after the list of corresponding authors. The
%% argument for \collaboration is the collaboration identifier. Authors are
%% encouraged to surround collaboration identifiers with ()s. The 
%% \nocollaboration command takes no argument and exists to indicate that
%% the nearby authors are not part of surrounding collaborations.

%% Mark off the abstract in the ``abstract'' environment. 
\begin{abstract}

Mid-Infrared (IR) silicate dust bands observed in heavily obscured active galactic nuclei (AGNs) include information on the mineralogical properties of silicate dust.
We aim to investigate the mineralogical picture of the circumnuclear region of heavily obscured AGNs to reveal obscured AGN activities through the picture.
In our previous study \citep{Tsuchikawa2021}, we investigated the properties of silicate dust in heavily obscured AGNs focusing on the mineralogical composition and the crystallinity with Spitzer/IRS 5.3-12$~\mu$m spectra.
In this study, we model the full-range Spitzer/IRS 5-30$~\mu$m spectra of 98 heavily obscured AGNs using a one-dimensional radiative transfer calculation with four dust species in order to evaluate wider ranges of the properties of silicate dust more reliably.
%In addition, we analyze the central wavelength of the local 23$~\mu$m crystalline feature.
Comparing fitting results between four dust models with a different size and porosity, 95 out of the 98 galaxies prefer a porous silicate dust model without micron-sized large grains. The pyroxene mass fraction and the crystallinity are overall consistent with but significantly different from the previous results for the individual galaxies. %because this study derives the dust properties more reliably by considering the radiative transfer effects.
%The peak position of the 23$~\mu$m crystalline feature varies significantly from 23 to 24$~\mu$m.
The pyroxene-poor composition, small dust size and high porosity are similar to newly formed dust around mass-loss stars as seen in our Galaxy, which presumably originates from the recent circumnuclear starburst activity. The high crystallinity on average suggests dust processing induced by AGN activities. 
%We conclude that the variations of the dust properties are likely attributed to the differences in the evolutionary scenario, the evolutionary stage and the viewing angle of a heavily obscured AGN. 
%abstract\footnote{Abstracts for Research Notes of the American Astronomical 
%Society (RNAAS) are limited to 150 words.  
%If you exceed this length the Editorial office will ask you to shorten it. This abstract has 182 words.

\end{abstract}

%% Keywords should appear after the \end{abstract} command. 
%% The AAS Journals now uses Unified Astronomy Thesaurus concepts:
%% https://astrothesaurus.org
%% You will be asked to selected these concepts during the submission process
%% but this old "keyword" functionality is maintained in case authors want
%% to include these concepts in their preprints.
%\keywords{Classical Novae (251) --- Ultraviolet astronomy(1736) --- History of astronomy(1868) --- Interdisciplinary astronomy(804)
\keywords{Active galactic nuclei (16) -- Infrared galaxies (790) -- Astrophysical dust processes (99) }

%% From the front matter, we move on to the body of the paper.
%% Sections are demarcated by \section and \subsection, respectively.
%% Observe the use of the LaTeX \label
%% command after the \subsection to give a symbolic KEY to the
%% subsection for cross-referencing in a \ref command.
%% You can use LaTeX's \ref and \label commands to keep track of
%% cross-references to sections, equations, tables, and figures.
%% That way, if you change the order of any elements, LaTeX will
%% automatically renumber them.
%%
%% We recommend that authors also use the natbib \citep
%% and \citet commands to identify citations.  The citations are
%% tied to the reference list via symbolic KEYs. The KEY corresponds
%% to the KEY in the \bibitem in the reference list below. 

\section{Introduction} \label{sec:intro}
An active galactic nucleus (AGN) experiences actively evolving phases of super-massive blackholes (SMBH) by releasing the gravitational energy of accreting materials. The presence of AGN is classically diagnosed using optical high-excitation lines \citep[e.g., ][]{Kauffmann2003}, which, however, reveals that the line diagnostics misses a significant fraction of AGNs due to extinction by large amounts of dust surrounding AGNs.
For example, more than 50\% of optically non-Seyfert ultra-luminous infrared galaxies (ULIRGs), whose infrared (IR) luminosity is defined to be higher than $10^{12}~L_{\odot}$, were reported to host heavily obscured AGNs on the basis of IR spectral analyses \citep{Imanishi2010, Ichikawa2014}. 
The morphological structure of U/LIRGs tends to show signature for galaxy interaction \citep{Sanders1996}.
Hydrodynamical simulations of the galaxy merger successfully reproduce various observed estimates such as the quasar luminosity function at each redshift \citep[e.g., ][]{Hopkins2006}. 
In these simulations, heavily obscured AGNs are predicted to be in an obscured phase of the nuclear growth before outflows blow out obscuring clouds, quench the star-forming activities and then evolve to unobscured AGNs.
Indeed, recent observations revealed the ubiquitous presence of outflows and/or inflows at the nuclear region in ULIRGs \citep[e.g., ][]{Toba2017, Veilleux2020}, suggesting dynamic evolutionary pictures of AGNs coupling with the surrounding material.

Silicate dust, which is a major component composing the cosmic dust, shows prominent spectral bands in the mid-IR wavelength range. The spectral features due to silicate dust have been detected in various kinds of astronomical objects such as AGNs \citep[e.g., ][]{Hao2007} as well as circumstellar disks or comets \citep[e.g., ][]{Molster2003, Henning2010}, through mid-IR spectroscopic observations with infrared space telescopes of ISO \citep{Kessler1996}, Spitzer \citep{Werner2004} and AKARI \citep{Murakami2007}.
Many papers focused on the spectral profiles of the silicate dust bands of circumstellar and cometary dust in detail, and discussed evolutionary scenarios of the system or astrophysical phenomena. In contrast, few studies systematically discussed the mid-IR spectra of heavily obscured AGNs for silicate dust properties in detail, although dust in heavily obscured AGNs is expected to be processed under more energetic and more dynamic environments than those surrounding the circumstellar or cometary dust.

\citet{Spoon2006} found crystallinity higher than 10\% for several ULIRGs. They concluded that the high crystallinity originates from starburst activities, because the mid-IR crystalline features are detected only in absorption and thus the crystalline silicate is likely to be located far from the hot nucleus. On the other hand, \citet{Kemper2011} suggested that the starburst activities alone cannot explain such a high crystallinity, who concluded that additional crystallization mechanisms are needed other than the mass ejection from mass-loss stars in the starburst activities.
\citet{Tsuchikawa2021} recently investigated the properties of silicate dust in heavily obscured AGNs using Spitzer/IRS archival data. They revealed that the crystallinity and the mineralogical olivine-to-pyroxene ratio are higher on average than those observed in the line of sight toward the diffuse interstellar medium (ISM) in our Galaxy. The olivine-rich mineralogical composition suggests that amorphous silicate is likely to be newly formed, which presumably originates from starburst activities. Moreover, on the basis of the above studies, \citet{Tsuchikawa2021} consider a scenario of dust processing in which amorphous silicate newly formed by starburst activities is crystallized in regions close to the nucleus by the AGN activities, and then transported to cooler regions by outflows.

It is important to investigate other properties of amorphous silicate dust, such as grain size and porosity, to discuss the dust processing scenario in more detail. As an example of silicate dust processing observed in our Galaxy, it is reported that amorphous silicate dust around YSOs is larger in size than that in the diffuse ISM, calling for the dust growth therein \citep[e.g., ][]{Juhasz2010}.  \citet{Tsuchikawa2021} conducted spectral fitting to the 10$~\mu$m absorption feature of the silicate dust bands, which did not constrain the size or porosity of amorphous silicate; the bottom profile of the 10$~\mu$m absorption feature is easily blurred by increases in the polycyclic aromatic hydrocarbon (PAH) or unobscured hot dust emission. 
Thus spectral fitting for the full wavelength range of 5--30$~\mu$m is important. A caveat of the full-range spectral fitting is the radiative transfer effect. Indeed, apparent optical depth ratios of the 10$~\mu$m to 18$~\mu$m amorphous silicate features of heavily obscured AGNs cannot be reproduced by the model in \citet{Tsuchikawa2021} assuming a simple full screen obscuration. \citet{Sirocky2008} reproduced the optical depth ratios of heavily obscured AGNs by performing a radiative transfer calculation, and constrained geometrical properties of the dust distribution or dust opacity models though the dust distribution in \citet{Sirocky2008} cannot approximate a clumpy AGN torus as done e.g., by \citet{Siebenmorgen2015}. %it is necessary to analyze the full-range mid-IR spectra considering the radiative transfer effects in order to accurately obtain the dust properties of the size and porosity as well as the crystallinity and mineralogical composition.
Hence, in the present study, we apply a model analysis to the full-range IRS spectra of heavily obscured AGNs including the wavelength range longer than 12$~\mu$m, which was outside the fitting range in the previous study, using a radiative transfer calculation to determine wider ranges of properties of silicate dust more reliably. 
On the basis of the dust properties thus obtained, we discuss the origin of each dust species or dust processing scenarios to imply a physical picture of the circumnuclear region of heavily obscured AGNs. Throughout the paper, we calculate the luminosity distance to the galaxies assuming the cosmological parameters $H_0 = 70~{\rm km~s^{-1}~Mpc^{-1}}$, $\Omega_{\Lambda}=0.7$ and $\Omega_{\rm m}=0.3$.

%------------------------------------------------------------------
%\section{Data and analysis}
\section{The sample\label{sec:select}} 
We selected the mid-IR spectra of nearby heavily obscured AGNs from the sample of \citet{Tsuchikawa2021}, which were observed by the low resolution mode of the InfraRed Spectrograph \citep[IRS;][]{Houck2004} onboard the Spitzer Space Telescope. 
The sample of the previous study was selected by the following three criteria: (1) the apparent optical depth of the 10$~\mu$m silicate feature larger than 1.5, (2) the equivalent width of the 6.2$~\mu$m PAH emission smaller than 270$~$nm and (3) the redshift lower than 0.35. The mid-IR spectra of the sample were retrieved from the Cornell AtlaS of Spitzer/IRS
Sources \citep[CASSIS;][]{Lebouteiller2011} version LR7 as done in \citet{Tsuchikawa2021}.
In this study, we analyzed the spectra in the full IRS spectral range. Because of the robustness of the analysis, we added a selection criterion that the apparent optical depth at 17$~\mu$m is larger than $0.2$. We adopted a power-law function used in \citet{Imanishi2009} for the absorption-free continuum, which was determined with the anchor points at 14.2 and 24$~\mu$m. By the additional criterion for the present study, 98 out of the 115 objects in the previous sample were selected.
We performed the spectral stitching as described in the previous study as well.
We did not use the bonus segments of the LL order (19.4--21.7$~\mu$m), because of mismatch between the first and second orders. %, which were out of scope of the main analysis in \citet{Tsuchikawa2021}

\startlongtable
\begin{deluxetable*}{lccccc}
\tablenum{1}
\tablecaption{General properties of the sample\label{tab:sample}}
\tabletypesize{\scriptsize}
\tablewidth{0pt}

\tablehead{
\colhead{Name} & \colhead{AORkey} & 
\colhead{R.A.~(J2000)} & \colhead{Dec.~(J2000)} & 
\colhead{$z$} & \colhead{log~$L_{\rm IR}$~[$\rm L_\odot$]}
} 
\decimalcolnumbers
\startdata
        IRAS~00091--0738 & 10440960, 10108928 & 00h11m43.2s & --07d22m06s & 0.1184 & 12.27~$\pm$~0.07 \\
       IRAS~F00183--7111 &            7556352 & 00h20m34.6s & --70d55m26s & 0.3270 & 12.95~$\pm$~0.09 \\
        IRAS~00188--0856 &            4962560 & 00h21m26.4s & --08d39m27s & 0.1284 & 12.41~$\pm$~0.09 \\
        IRAS~00397--1312 &            4963584 & 00h42m15.4s & --12d56m03s & 0.2617 & 12.94~$\pm$~0.19 \\
        IRAS~00406--3127 &            4964096 & 00h43m03.1s & --31d10m49s & 0.3424 & 12.80~$\pm$~0.15 \\
      IRAS~01166--0844SE & 10441984, 10109952 & 01h19m07.8s & --08d29m12s & 0.1180 & 12.12~$\pm$~0.13 \\
        IRAS~01199--2307 &            4964864 & 01h22m20.8s & --22d51m57s & 0.1562 & 12.31~$\pm$~0.13 \\
        IRAS~01298--0744 &            4965120 & 01h32m21.4s & --07d29m08s & 0.1362 & 12.36~$\pm$~0.12 \\
        IRAS~01355--1814 &            4965376 & 01h37m57.4s & --17d59m20s & 0.1920 & 12.48~$\pm$~0.11 \\
        IRAS~F01478+1254 &           23012864 & 01h50m28.4s &  +13d08m58s & 0.1470 & 11.98~$\pm$~0.39 \\
        IRAS~01569--2939 &           10110208 & 01h59m13.7s & --29d24m34s & 0.1400 & 12.26~$\pm$~0.11 \\
        IRAS~02455--2220 &            4967680 & 02h47m51.2s & --22d07m38s & 0.2840 & 12.70~$\pm$~0.15 \\
         IRAS~02530+0211 &            6652160 & 02h55m34.4s &  +02d23m41s & 0.0276 & 11.05~$\pm$~0.05 \\
         IRAS~03158+4227 &           12256256 & 03h19m11.9s &  +42d38m25s & 0.1344 & 12.61~$\pm$~0.08 \\
        NGC~1377\tablenotemark{\tiny a} &            9511424 & 03h36m40.1s & --20d54m02s & 0.0060 & 10.17~$\pm$~0.03 \\
        IRAS~03538--6432 &            4968192 & 03h54m25.2s & --64d23m44s & 0.3007 & 12.79~$\pm$~0.10 \\
         IRAS~03582+6012 &           20341504 & 04h02m32.9s &  +60d20m41s & 0.0300 & 11.40~$\pm$~0.09 \\
        IRAS~04074--2801 &           25185536 & 04h09m30.4s & --27d53m43s & 0.1537 & 12.25~$\pm$~0.11 \\
        IRAS~04313--1649 &            4968960 & 04h33m37.0s & --16d43m31s & 0.2680 & 12.67~$\pm$~0.11 \\
        IRAS~04384--4848 &            6650880 & 04h39m50.8s & --48d43m17s & 0.2035 & 12.40~$\pm$~0.07 \\
          ESO~203--IG001 &           20334080 & 04h46m49.5s & --48d33m30s & 0.0529 & 11.85~$\pm$~0.04 \\
        IRAS~05020--2941 &           25185792 & 05h04m00.7s & --29d36m54s & 0.1544 & 12.38~$\pm$~0.06 \\
       IRAS~F06076--2139 &           20359680 & 06h09m45.7s & --21d40m24s & 0.0374 & 11.63~$\pm$~0.04 \\
        IRAS~06206--6315 &            4969984 & 06h21m00.8s & --63d17m23s & 0.0924 & 12.22~$\pm$~0.04 \\
        IRAS~06301--7934 &            4970240 & 06h26m42.2s & --79d36m30s & 0.1564 & 12.39~$\pm$~0.06 \\
        IRAS~06361--6217 &            4970496 & 06h36m35.7s & --62d20m31s & 0.1596 & 12.41~$\pm$~0.10 \\
        IRAS~07251--0248 &           20346112 & 07h27m37.6s & --02d54m54s & 0.0876 & 12.41~$\pm$~0.08 \\
         IRAS~08201+2801 &           18202112 & 08h23m12.6s &  +27d51m40s & 0.1678 & 12.30~$\pm$~0.14 \\
       IRAS~F08520--6850 &           20343808 & 08h52m32.0s & --69d01m54s & 0.0451 & 11.74~$\pm$~0.04 \\
         IRAS~08572+3915 &            4972032 & 09h00m25.3s &  +39d03m54s & 0.0584 & 12.15~$\pm$~0.03 \\
         IRAS~09539+0857 & 10444032, 11676160 & 09h56m34.3s &  +08d43m05s & 0.1289 & 12.10~$\pm$~0.19 \\
       IRAS~F10038--3338 &           20352256 & 10h06m04.6s & --33d53m06s & 0.0342 & 11.71~$\pm$~0.05 \\
         IRAS~10091+4704 &            4973824 & 10h12m16.7s &  +46d49m42s & 0.2460 & 12.65~$\pm$~0.10 \\
         IRAS~10173+0828 & 14838528, 20314880 & 10h20m00.2s &  +08d13m34s & 0.0491 & 11.80~$\pm$~0.12 \\
        IRAS~F10237+4720 &           22117632 & 10h26m48.2s &  +47d05m07s & 0.0589 & 11.48~$\pm$~0.11 \\
         IRAS~10378+1109 &            4974336 & 10h40m29.1s &  +10d53m17s & 0.1363 & 12.35~$\pm$~0.09 \\
        IRAS~10485--1447 & 10444800, 10105088 & 10h51m03.0s & --15d03m22s & 0.1330 & 12.22~$\pm$~0.16 \\
         IRAS~11028+3130 &           18203392 & 11h05m37.5s &  +31d14m31s & 0.1986 & 12.42~$\pm$~0.14 \\
         IRAS~11038+3217 &            4975104 & 11h06m35.7s &  +32d01m46s & 0.1300 & 11.62~$\pm$~0.35 \\
        IRAS~11095--0238 &            4975360 & 11h12m03.3s & --02d54m24s & 0.1066 & 12.28~$\pm$~0.08 \\
        IRAS~11130--2659 &           10105600 & 11h15m31.5s & --27d16m22s & 0.1361 & 12.14~$\pm$~0.15 \\
         IRAS~11180+1623 &           18203648 & 11h20m41.7s &  +16d06m56s & 0.1660 & 12.31~$\pm$~0.14 \\
        IRAS~11223--1244 &            4976128 & 11h24m50.7s & --13d01m16s & 0.1990 & 12.57~$\pm$~0.10 \\
         IRAS~11506+1331 & 10445312, 10111488 & 11h53m14.1s &  +13d14m26s & 0.1273 & 12.35~$\pm$~0.10 \\
         IRAS~11524+1058 &           18203904 & 11h55m05.1s &  +10d41m22s & 0.1787 & 12.23~$\pm$~0.15 \\
         IRAS~11582+3020 &            4976384 & 12h00m46.8s &  +30d04m14s & 0.2230 & 12.57~$\pm$~0.17 \\
         IRAS~12032+1707 &            4976896 & 12h05m47.7s &  +16d51m08s & 0.2178 & 12.63~$\pm$~0.17 \\
        IRAS~12127--1412 & 10445824, 10105856 & 12h15m19.1s & --14d29m41s & 0.1330 & 12.20~$\pm$~0.13 \\
       IRAS~F12224--0624 &           20367104 & 12h25m03.9s & --06d40m52s & 0.0264 & 11.24~$\pm$~0.08 \\
                NGC~4418 &            4935168 & 12h26m54.6s &  -00d52m40s & 0.0073 & 11.04~$\pm$~0.05 \\
        IRAS~12359--0725 &           10106112 & 12h38m31.6s & --07d42m25s & 0.1380 & 12.18~$\pm$~0.22 \\
         IRAS~12447+3721 &           25187840 & 12h47m07.7s &  +37d05m36s & 0.1580 & 12.17~$\pm$~0.20 \\
        IRAS~F13045+2354 &            4168448 & 13h07m00.6s &  +23d38m04s & 0.2750 & 12.61~$\pm$~0.23 \\
        IRAS~13106--0922 &           25186048 & 13h13m14.6s & --09d38m08s & 0.1745 & 12.57~$\pm$~0.22 \\
        IRAS~F13279+3401 &           12235264 & 13h30m15.2s &  +33d46m29s & 0.0230 & 10.46~$\pm$~0.15 \\
         IRAS~13352+6402 &            4979968 & 13h36m51.1s &  +63d47m04s & 0.2366 & 12.55~$\pm$~0.10 \\
                 Mrk~273 &            4980224 & 13h44m42.1s &  +55d53m13s & 0.0378 & 12.15~$\pm$~0.03 \\
         IRAS~14070+0525 &            4980992 & 14h09m31.2s &  +05d11m31s & 0.2644 & 12.82~$\pm$~0.12 \\
        IRAS~F14394+5332 &           29040128 & 14h41m04.3s &  +53d20m08s & 0.1045 & 12.10~$\pm$~0.06 \\
        IRAS~F14511+1406 &            4168960 & 14h53m31.5s &  +13d53m58s & 0.1390 & 11.93~$\pm$~0.20 \\
        IRAS~F14554+3858 &           28244224 & 14h57m22.7s &  +38d46m28s & 0.0735 & 11.10~$\pm$~0.29 \\
         IRAS~15225+2350 &           10112512 & 15h24m43.9s &  +23d40m10s & 0.1390 & 12.17~$\pm$~0.09 \\
         IRAS~15250+3609 &            4983040 & 15h26m59.3s &  +35d58m37s & 0.0552 & 12.05~$\pm$~0.05 \\
                 Arp~220 &            4983808 & 15h34m57.2s &  +23d30m11s & 0.0181 & 12.17~$\pm$~0.03 \\
FESS~J160655.82+541500.7 &           24189952 & 16h06m55.8s &  +54d15m00s & 0.2060 &    --\tablenotemark{\tiny c} \\
        IRAS~F16073+0209 &           17546496 & 16h09m49.7s &  +02d01m30s & 0.2230 & 12.35~$\pm$~0.27 \\
        IRAS~16090--0139 &            4984576 & 16h11m40.4s & --01d47m05s & 0.1336 & 12.57~$\pm$~0.04 \\
        IRAS~F16156+0146 &           17546752 & 16h18m09.3s &  +01d39m22s & 0.1320 & 12.11~$\pm$~0.13 \\
        IRAS~F16242+2218 &           17547008 & 16h26m26.0s &  +22d11m45s & 0.1570 & 11.74~$\pm$~0.29 \\
        IRAS~F16305+4823 &           22135040 & 16h31m58.7s &  +48d17m22s & 0.0874 & 11.92~$\pm$~0.08 \\
         IRAS~16300+1558 &            4985088 & 16h32m21.4s &  +15d51m45s & 0.2417 & 12.74~$\pm$~0.11 \\
         IRAS~16455+4553 &           14875136 & 16h46m58.9s &  +45d48m22s & 0.1906 & 12.37~$\pm$~0.09 \\
        IRAS~16468+5200W &           10107136 & 16h48m01.3s &  +51d55m43s & 0.1500 & 12.11~$\pm$~0.11 \\
        IRAS~16468+5200E &           10106880 & 16h48m01.6s &  +51d55m44s & 0.1500 & 12.11~$\pm$~0.11 \\
         IRAS~17044+6720 &           10107904 & 17h04m28.4s &  +67d16m28s & 0.1349 & 12.17~$\pm$~0.08 \\
        IRAS~F17028+3616 &           27194112 & 17h04m33.5s &  +36d12m18s & 0.0851 & 11.15~$\pm$~0.43 \\
         IRAS~17068+4027 &            4986112 & 17h08m32.1s &  +40d23m28s & 0.1790 & 12.40~$\pm$~0.10 \\
        IRAS~17208--0014 &            4986624 & 17h23m21.9s &  -00d17m00s & 0.0428 & 12.40~$\pm$~0.04 \\
         IRAS~17463+5806 &            4987392 & 17h47m04.7s &  +58d05m22s & 0.3090 & 12.64~$\pm$~0.11 \\
         IRAS~17540+2935 &           18204928 & 17h55m56.1s &  +29d35m26s & 0.1081 & 11.87~$\pm$~0.09 \\
         IRAS~18443+7433 &            4987904 & 18h42m54.7s &  +74d36m21s & 0.1347 & 12.32~$\pm$~0.08 \\
        IRAS~18531--4616 &            4988160 & 18h56m53.0s & --46d12m46s & 0.1408 & 12.33~$\pm$~0.22 \\
         IRAS~18588+3517 &           18205440 & 19h00m41.1s &  +35d21m27s & 0.1067 & 11.97~$\pm$~0.10 \\
        IRAS~20100--4156 &            4989696 & 20h13m29.8s & --41d47m34s & 0.1296 & 12.64~$\pm$~0.06 \\
        IRAS~20109--3003 &           14875904 & 20h14m05.5s & --29d53m53s & 0.1407 & 11.98~$\pm$~0.26 \\
         IRAS~20286+1846 &           18205696 & 20h30m54.4s &  +18d56m37s & 0.1358 & 12.20~$\pm$~0.26 \\
        IRAS~20551--4250 &            4990208 & 20h58m26.7s & --42d39m01s & 0.0430 & 12.06~$\pm$~0.03 \\
         IRAS~21077+3358 &           18205952 & 21h09m50.6s &  +34d10m34s & 0.1767 & 12.41~$\pm$~0.45 \\
         IRAS~21272+2514 &            4990464 & 21h29m29.3s &  +25d27m55s & 0.1508 & 12.30~$\pm$~0.39 \\
       IRAS~F21329--2346 & 10448640, 10108160 & 21h35m45.8s & --23d32m34s & 0.1251 & 12.15~$\pm$~0.10 \\
       IRAS~22088--1831W &           25189120 & 22h11m33.7s & --18d17m06s & 0.1702 & 12.44~$\pm$~0.12 \\
       IRAS~22088--1831E &           25189376 & 22h11m33.8s & --18d17m05s & 0.1702 & 12.44~$\pm$~0.12 \\
         IRAS~22116+0437 &           18206464 & 22h14m10.3s &  +04d52m26s & 0.1938 & 12.33~$\pm$~0.27 \\
        NGC~7479\tablenotemark{\tiny b} &           22093312 & 23h04m56.6s &  +12d19m22s & 0.0079 & 10.76~$\pm$~0.06 \\
         IRAS~23129+2548 &            4991488 & 23h15m21.4s &  +26d04m32s & 0.1789 & 12.48~$\pm$~0.12 \\
        IRAS~F23234+0946 & 10449152, 10108416 & 23h25m56.2s &  +10d02m50s & 0.1279 & 12.15~$\pm$~0.11 \\
        IRAS~23230--6926 &            4992000 & 23h26m03.5s & --69d10m20s & 0.1066 & 12.31~$\pm$~0.04 \\
         IRAS~23365+3604 &            4992512 & 23h39m01.2s &  +36d21m09s & 0.0645 & 12.17~$\pm$~0.06 \\
\enddata

\tablecomments{Column 1: the name of the object; Column 2: AORkey (Spitzer/IRS identification number); Columns 3, 4: the position of the object; Column 5: the redshift cited from the NASA/IPAC Extragalactic Database (NED); Column 6: the total 8--1000$~\mu$m IR luminosity, which is calculated from the IRAS fluxes with the definition by \citet{Sanders1996}.}
\tablenotetext{\tiny a}{The spectrum of NGC~1377 is unavailable in CASSIS. The spectral data were retrieved from the summary of the SINGS Legacy project \citep{Kennicutt2003, SINGS} in the NASA/IPAC IR Science Archive (IRSA).} 
\tablenotetext{\tiny b}{The SL order 2 spectrum of NGC~7479 is unavailable in CASSIS. It was retrieved from the Spitzer Heritage Archive (SHA).}
\tablenotetext{\tiny c}{The IR fluxes cannot be obtained. Thus its IR luminosity cannot be calculated.}

\end{deluxetable*}

\section{Modeling of the full-range IRS spectra \label{sec:model}}
\begin{figure*}
        \centering
        \includegraphics[width=16cm,clip]{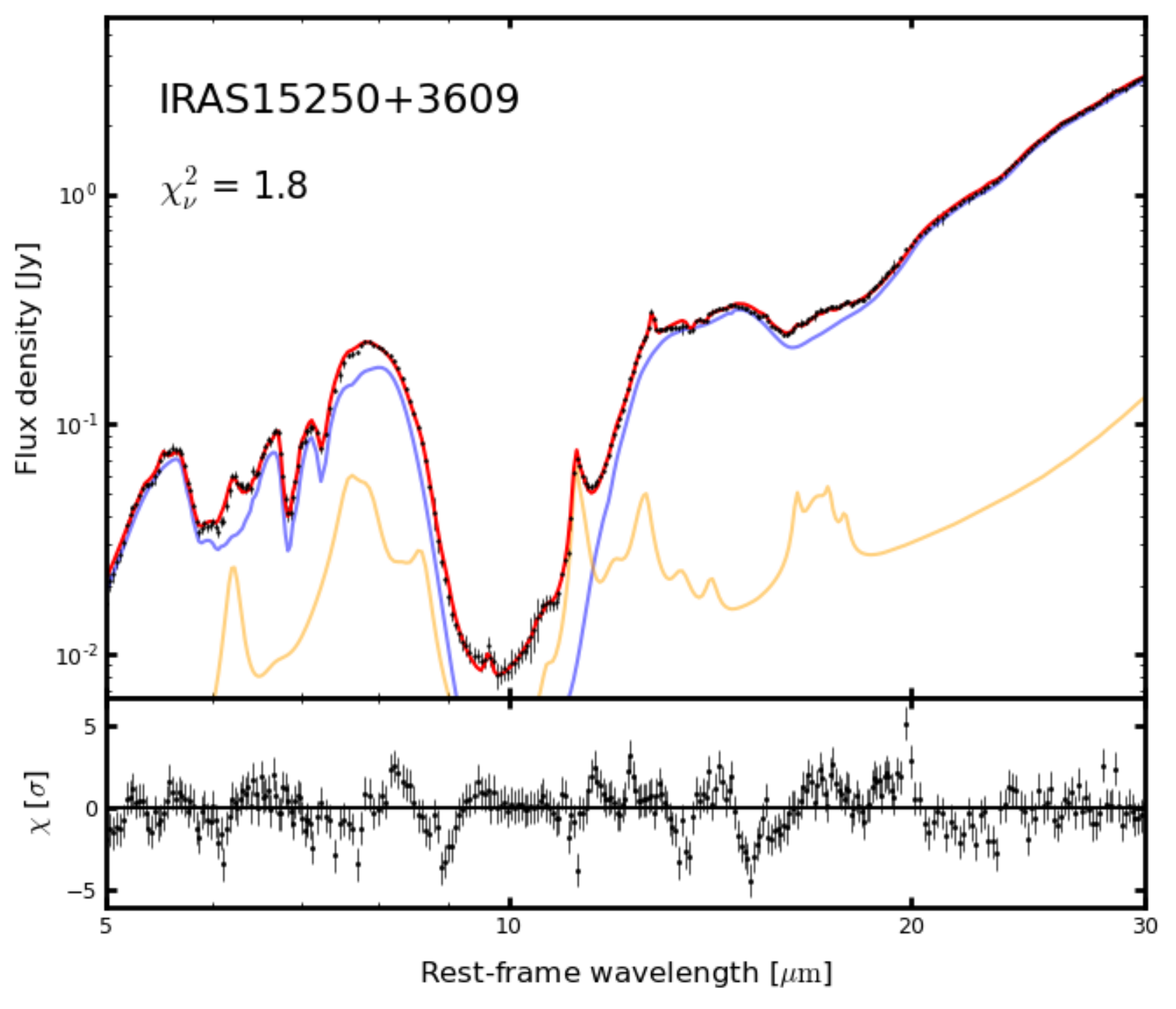}
        \caption{An example result of the mid-IR 5--30$~\mu$m spectral modeling of heavily obscured AGNs. The black points and the red solid line show the observed spectrum and the best-fit model of IRAS~15250+3609. The blue solid line represents the AGN-heated dust emission components with ice absorption, $F_{\rm d}^{\rm agn} {\rm exp}(-\tau_{\rm ice})$, while the yellow lines represent the SF-heated dust and PAH emission, $F_{\rm sf}$, and the line emission components. %The gray dashed line shows the emission component due to the unobscured hot dust heated by AGN.
        We show the reduced $\chi^2$ value in the upper left corner. The residual spectrum normalized by the errors is depicted in the bottom panel.}
        \label{fig:typical}%
   \end{figure*}

%\citet{Tsuchikawa2021} conducted the spectral fitting within the wavelength range of 5.3--12$~\mu$m. To obtain more information in the dust properties, we expand the modeling range to 5--30$~\mu$m.
For the purpose of modeling the full-range IRS spectra, it is important to determine the mid-IR continuum shape.
The continuum emission is likely to originate from the dust heated by AGN. \citet{Tsuchikawa2021} applied a power-law plus spline function simply assuming a full screen obscuration by silicate dust to reproduce the AGN-heated dust continuum emission. Although the model used in \citet{Tsuchikawa2021} is a good approximation for the narrow wavelength range of 5.3--12$~\mu$m, it cannot well reproduce the full-range 5--30$~\mu$m spectra in our sample. The full screen dust obscuration model overpredicts the apparent optical depth of the 18$~\mu$m silicate feature compared to the 10$~\mu$m silicate feature. As described in Sect.~\ref{sec:intro}, this is likely due to a radiative transfer effect, as the effective absorption becomes shallower at longer wavelengths, assuming that the continuum source at longer wavelengths is located farther from the nucleus \citep[e.g.,][]{Spoon2006, Imanishi2006, Sirocky2008}. The spectral modelings of the full-range IRS spectra of heavily obscured AGNs were performed by several papers. \citet{Marshall2007} and \citet{Stierwalt2013,Stierwalt2014} reproduced the IRS spectra using multi-component dust emission models with extinction curves of our Galaxy. On the other hand, for instance, \citet{Siebenmorgen2007} modeled the spectra by a radiative transfer calculation, considering realistic geometry of the dust distribution. However, these studies did not focus on the optical properties of silicate dust in detail, and thus could not obtain good fits for the silicate features as a whole. 
In this study, we achieve considerably good fits to the full-range IRS spectra in our sample, as shown in Fig.~\ref{fig:typical}, by modeling dust properties in detail and numerically calculating the radiative transfer of dusty shells. The radiative transfer modeling of the AGN-heated dust emission, $F_{\rm d}^{\rm agn}$, is explained in the following subsections.

%Hence, in this study, we obtain the model spectrum of the AGN-heated dust emission, $F_{\rm d}^{\rm agn}$, by a numerical calculation of the radiative transfer assuming dusty shells as the geometrical distribution of the dust and various dust properties, the detailed explanations of which are described in the following subsections.
%considerably well

% \citet{Marshall2007}, \citet{Veilleux2009} and \citet{Stierwalt2013,Stierwalt2014} reproduce the IRS spectra using multi-component dust emission models and the extinction curves in our Galaxy or which is empirically derived.

We cannot reproduce the sample spectra by the component of $F_{\rm d}^{\rm agn}$ alone; a significant fraction of the sample spectra clearly show the 6--8$~\mu$m absorption features due to $\rm H_2O$ ice and hydrogenated amorphous carbon (HAC). 
We reproduced these features using the 3 templates derived in \citet{Tsuchikawa2021} and Gaussian functions by multiplying the extinction term of ${\rm exp}(-\tau_{\rm ice})$ to $F_{\rm d}^{\rm agn}$, the parameters of which were fixed at the same values as those in \citet{Tsuchikawa2021}.
%Thus, we assumed a picture that the dust continuum $F_{\rm d}^{\rm agn}$ is completely obscured by the ice and HAC located in the foreground cooler region, and  
Furthermore, we newly added two broad absorption components at $\sim$4.7 and 12$~\mu$m, which are most probably due to CO gas ro-vibrational and $\rm H_2O$ ice libration modes, respectively. For the opacity profiles of the 4.7 and 12$~\mu$m absorptions, we used a Gaussian function and an absorption coefficient of a 0.1~$\mu$m-sized homogeneous sphere of 10~K pure $\rm H_2O$ ice, respectively. The central wavelength and full width at half maximum (FWHM) of the Gaussian function for the 4.7$~\mu$m feature were fixed at 4.8$~\mu$m and 0.75$~\mu$m, respectively. The absorption coefficient for the 12$~\mu$m feature was calculated by the Mie theory using the optical constants measured by \citet{Hudgins1993}.
We also took into account an unobscured hot dust emission heated by AGN in order to model the flat profile at the bottom of the 10$~\mu$m silicate absorption feature as mentioned in \citet{Tsuchikawa2021}. 
We adopted the average spectrum of quasars obtained by \citet{Hao2007} for the spectral profile of the AGN-heated unobscured hot dust component.

%\startlongtable
\begin{deluxetable*}{clcc}
\tablenum{2}
\tablecaption{Free parameters for the mid-IR spectral modeling\label{tab:param}}
\tabletypesize{\scriptsize} 
\tablewidth{0pt}
\tablehead{
\colhead{Parameter} & \colhead{Description} & 
\colhead{Range} & \colhead{Grid for calculation with DUSTY} 
} 
\startdata   
		\multicolumn{2}{l}{\multirow{2}{*}{AGN-heated dust emission}}\\\\
	    $C_{\rm d}^{\rm agn}$ & Amplitude of $F_{\rm d}^{\rm agn}$  & [0, $\infty$] & -- \\
	    $f_{\rm BB}$ & Fractional contribution of the blackbody of 1000~K  & [0, $\infty$] & -- \\
	    $\tau_{\rm lib,~peak}$ & Amplitude of the absorption due to the $\rm H_2O$ ice libration mode  & [0, $\infty$] & -- \\
	    $C^{\rm agn}_{\rm unobs}$ & Amplitude of the unobscured AGN-heated dust emission  & [0, $\infty$] & -- \\
		\multicolumn{2}{l}{\multirow{2}{*}{SF-heated dust and PAH emission}}\\\\ 
	    $C_{\rm sf}$ & Amplitude of the SF-heated dust and PAH emission  & [0, $\infty$] & -- \\
	    $f_{\rm cont/7}$ & SF-heated dust to PAH 7.7~$\mu$m emission ratio  & [0.3, 1.5] & -- \\
	    $f_{6/7}$ & PAH 6.2-to-7.7~$\mu$m emission ratio  & [0.7, 1.5] & -- \\
	    $f_{11/7}$ & PAH 11.2-to-7.7~$\mu$m emission ratio  & [0.5, 3] & -- \\
	    $f_{17/7}$ & PAH 17-to-7.7~$\mu$m emission ratio  & [0.5, 2] & -- \\
	    $\tau_{\rm sf, 9.7}$ & Peak dust extinction for the SF-heated dust and PAH emissions  & [0.3, 1.5] & -- \\\\
		\tableline
		\multicolumn{2}{l}{\multirow{2}{*}{Dust composition ratios}} \\\\
		$r_{\rm py}$ & Mass ratio of amorphous pyroxene to total amorphous silicate & [0, 40]\% & 0, 10, 20, 30, 40\\
		$r_{\rm cr}$ & Mass ratio of crystalline to total silicate & [0, 20]\% & 0, 5, 10, 15, 20\\
		$r_{\rm si}$ & Mass ratio of silicate to total dust & [60, 97]\% & 60, 65, 70, 75, 80, 85, 90, 95, 97\\
		\multicolumn{2}{l}{\multirow{2}{*}{Geometrical properties}}\\ \\
		$p$ & Power-law index of the radial density profile & [0.5, 2.0] & 0.5, 1.0, 1.5, 2.0\\
		$Y$ & Ratio of the outer to inner radius of the dust shell & log [25, 800] & 25, 50, 100, 200, 400, 800\\
		\multirow{4}{*}{$\tau_{V}$} & \multirow{4}{*}{Dust optical depth at the wavelength of 550~nm} & \multirow{4}{*}{[10, 400]} &         10, 20, 30, 40, 50, 60, 70, 80,\\
		 & & &  90, 100, 110, 120, 130, 140, 150,\\
		 & & &  160, 170, 180, 190, 200, 210,\\
		 & & &  220, 230, 240, 260, 300, 350, 400\\
\enddata
\end{deluxetable*}

It is needed to consider an emission component due to the PAH and dust heated by SF activity, $F_{\rm sf}$, as well. 
In general, the inter-band ratios of the PAH emission vary among galaxies. For example, the strength ratio of the C-C stretching at 5--9$~\mu$m to C-H bending at 9--15$~\mu$m features is known to depend on the ionization degree of the PAH molecules \citep{Draine2007}. The average size of the PAHs also affects the inter-band ratios. \citet{Maaskant2014} show that AGN spectra tend to have relatively high ratios of the 17$~\mu$m complex to the other shorter wavelength features, which indicates that large PAHs are abundant in AGNs.
Moreover, for partially-extended sources, the ratio of the PAH emission to the AGN-heated dust emission is expected to be higher in LL spectra than in SL spectra because the slit size of the LL module is larger than that of the SL module.

\citet{Marshall2007} modeled the dust and PAH emission spectra typical of starburst galaxies with the templates derived from the spectra of NGC~7714 and the average spectra of the starburst galaxies in \citet{Brandl2006}, respectively.
They determined the relative strengths of the individual PAH emission bands using PAHFIT \citep{Smith2007} which can decompose the PAH emission component of a mid-IR spectrum with a multi-component Drude function.
In this study, we adopted the models by \citet{Marshall2007} for the SF-heated dust and PAH emission models.
In order to perform spectral fitting with a moderate degree of freedom, we decomposed the PAH model by \citet{Marshall2007} into the four components of $F_{\rm pah6}$, $F_{\rm pah7}$, $F_{\rm pah11}$ and $F_{\rm pah17}$ according to the central wavelengths of the PAH bands described by the Drude profiles as follows: 
\begin{equation}
    F_{\rm sf} = C_{\rm sf}~(f_{\rm dust/7}F_{\rm d}^{\rm sf} + f_{6/7}F_{\rm pah6} + F_{\rm pah7} + f_{11/7}F_{\rm pah11} + f_{17/7}F_{\rm pah17})~\frac{1-{\rm exp}(-\tau_{\rm sf})}{\tau_{\rm sf}},
\end{equation}
where $C_{\rm sf}$ is the amplitude of $F_{\rm sf}$, and $f_{\rm dust/7}$, $f_{6/7}$, $f_{11/7}$ and $f_{17/7}$ are the amplitude ratios of the individual components to $F_{\rm pah7}$. 
The individual four PAH components of $F_{\rm pah6}$, $F_{\rm pah7}$, $F_{\rm pah11}$ and $F_{\rm pah17}$ consist of multi-component Drude functions whose central wavelengths are in ranges of 5--7$~\mu$m, 7--9$~\mu$m, 11--15$~\mu$m and longer than 15$~\mu$m, respectively. Within the individual four PAH components, the inter-band ratios of the Drude functions were fixed.
$f_{\rm dust/7}$, $f_{6/7}$, $f_{11/7}$ and $f_{17/7}$ were constrained within plausible ranges on the basis of the variations of the PAH features in the starburst galaxies studied by \citet{Smith2007} and \citet{Draine2021b}, which are summarized in Table~\ref{tab:param}. 
The term of $(1-{\rm exp}(-\tau_{\rm sf}))/\tau_{\rm sf}$ reproduces the extinction for the SF-heated dust and PAH emission assuming a well-mixed geometry. The extinction curve observed in the line of sight toward the Galactic center \citep{Chiar2006} was used for the wavelength dependence of the optical depth, $\tau_{\rm sf}$. 
\citet{Tsuchikawa2021} considered an additional feature at 10.68$~\mu$m, which is likely due to dehydrogenated PAHs \citep{Mackie2015}. We incorporated it into the PAH model with the strength ratio of $F_{\rm pah11}$ to the 10.68$~\mu$m feature fixed at the value obtained from the average spectra of the starburst galaxies in \citet{Brandl2006}. 
In addition, both of the AGN-originated and SF-originated atomic and molecular line emissions were reproduced by Gaussian functions in the same way as in \citet{Tsuchikawa2021}.

\subsection{Radiative transfer calculation}
    
 We calculated the AGN-heated dust emission, $F_{\rm d}^{\rm agn}$, using the one-dimensional dust radiative transfer code DUSTY \citep{Ivezic1997}. DUSTY performs a self-consistent radiative transfer simulation, assuming a simple spatial distribution and the thermal equilibrium of dust. We assume a spherically symmetric dust distribution. Such a simplified geometry cannot approximate a clumpy AGN torus \citep[e.g.,][]{Siebenmorgen2015}. A power-law function was adopted for the radial mass density profile using the power-law index $p$, as described by $\rho(r) = \rho_0 r^{-p}$.
    The radial mass density was normalized by the radial optical depth at the optical wavelength of 0.55~$\mu$m, $\tau_{\rm V}=(\kappa_{\rm abs,V}+\kappa_{\rm sca,V})\rho_0{\int}^{r_{\rm out}}_{r_{\rm in}}r^{-p}dr$, where the mass absorption and scattering coefficients of dust, $\kappa_{\rm abs,\lambda}$ and $\kappa_{\rm sca,\lambda}$, are assumed to be spatially uniform. The wavelength dependences of $\kappa_{\rm abs,\lambda}$ and $\kappa_{\rm sca,\lambda}$, which include information on dust properties, are described in the next subsection.
    The inner radius of the spherically symmetric dust distribution, $r_{\rm in}$, is determined by the dust sublimation.  While the sublimation temperature of dust depends on dust species, DUSTY does not allow us to set different sublimation temperatures between the dust species. 
    Therefore we simply assumed 1000~K as the sublimation temperature of all the dust species. The outer radius, $r_{\rm out}$, is defined using the ratio of the outer to inner radii of dust cloud, $Y$, as $Y=r_{\rm out}/r_{\rm in}$.
    We adopted the spectrum of an accretion disk empirically derived by \citet{Schartmann2005} for the heating source buried in dust cloud. The source spectrum consists of a broken power-law function for the wavelength range of 0.01--10$~\mu$m and a 1000~K Planck function for wavelengths longer than 10$~\mu$m. 
    
      We tentatively fitted the sample spectra by the model described above to find that the above model often under-estimates the 5--8~${\mu}$m continuum of the sample spectra. 
    %These spectra typically have the flat continuum at wavelengths shorter than 7$~\mu$m compared to the steep continuum at wavelengths longer than 20$~\mu$m. 
    For a significant fraction of our sample AGNs, obscured outflows were detected \citep{Veilleux2020}. Hence we need to consider the contribution from the shock heating or the obscured polar dust emission, and thus introduce another obscured hot dust component using a Planck function with a fixed temperature of 1000~K. The hot dust component is assumed to be obscured by the dust shell considered in the DUSTY calculation, as described by,
\begin{equation}
    F_{\rm d}^{\rm agn} = C_{\rm d}^{\rm agn}~(F_{\rm DUSTY} + f_{\rm BB}B_{\rm \nu}(1000~{\rm K})~e^{-{\tau}_{{\rm ext}}}),
\label{eq:fagn}
\end{equation}
     where $F_{\rm DUSTY}$ and $B_{\rm \nu}(1000~{\rm K})$ correspond to the output spectrum of the DUSTY calculation and the blackbody emission spectrum with 1000~K, respectively. $C_{\rm d}^{\rm agn}$, $f_{\rm BB}$ and ${\tau}_{{\rm ext}}$ are the amplitude of $F_{\rm d}^{\rm agn}$, the fractional contribution of the blackbody component and the radial optical depth due to dust extinction, respectively.

\subsection{Dust properties}
\begin{figure*}
        \centering
        \includegraphics[width=14cm,clip]{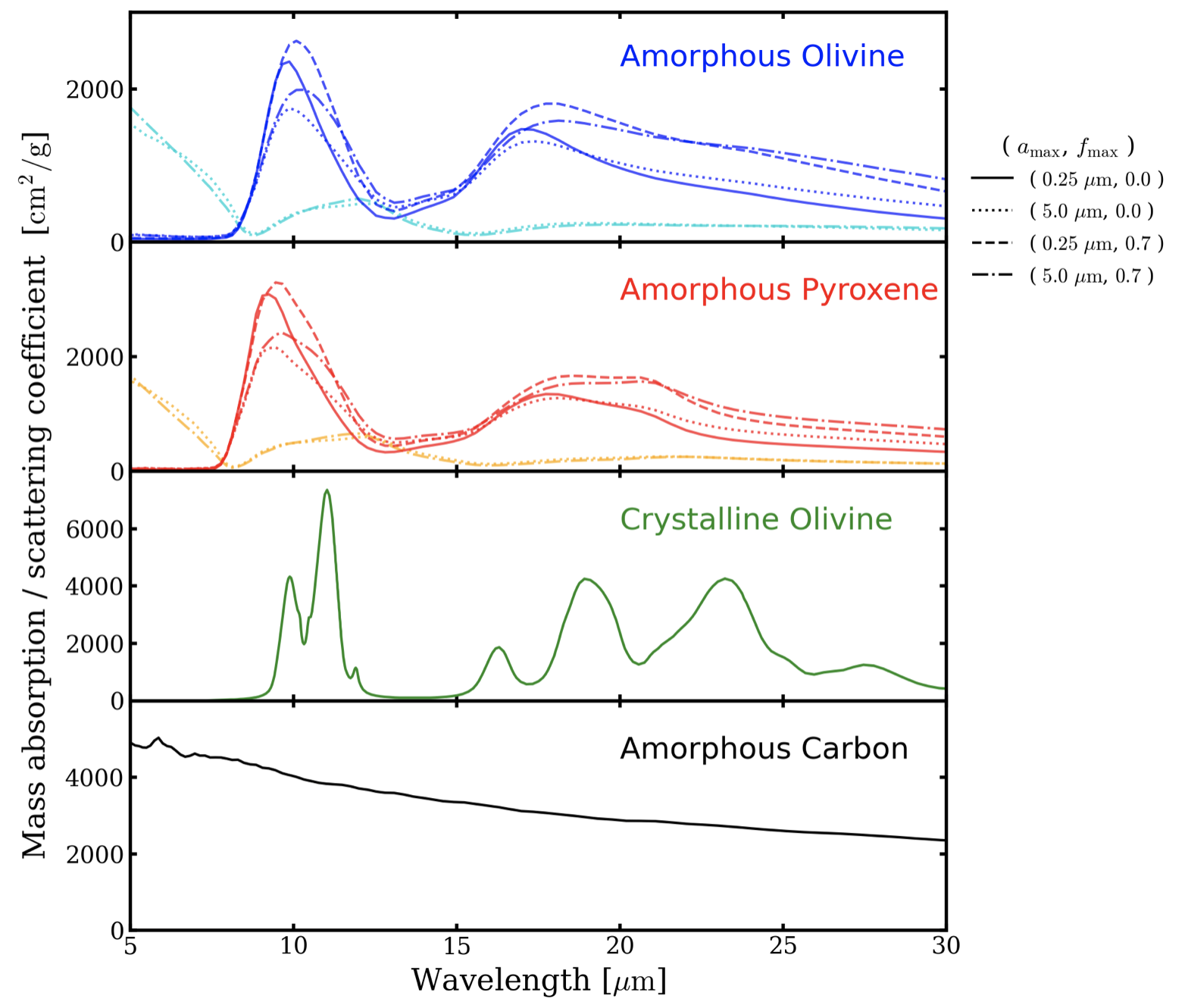}
        \caption{Mass absorption and scattering coefficients of amorphous olivine (top), amorphous pyroxene (upper middle), crystalline olivine (lower middle) and amorphous carbon (bottom). In the panels of amorphous olivine and pyroxene, the mass absorption and scattering coefficients calculated with different $f_{\rm max}$ and $a_{\rm max}$ are shown with different line styles. The cyan and orange lines show the mass scattering coefficients, and the other colors show the mass absorption coefficients. The mass scattering coefficients of amorphous olivine and pyroxene calculated with $a_{\rm max}=0.25~{\rm {\mu}m}$, crystalline olivine and amorphous carbon are much smaller than their mass absorption coefficients, and therefore not visible in the figure.}
        \label{fig:dust}%
   \end{figure*}

We constructed dust opacity models with the optical properties of amorphous olivine, amorphous pyroxene, crystalline olivine and amorphous carbon which were obtained by the experimental measurements, as summarized in Table~\ref{tab:dust} and Fig.~\ref{fig:dust}. 
As shown in the upper panels of Fig.~\ref{fig:dust}, amorphous olivine and pyroxene have different peak wavelengths of $\sim9.7$ and $\sim9.3~\mu$m, respectively. \citet{Tsuchikawa2019} found difference in the wing of the 10$~\mu$m silicate feature of heavily obscured AGNs at shorter wavelengths. Thus we incorporated amorphous olivine and pyroxene into the dust model for the purpose of reproducing the difference in the wing of the 10$~\mu$m feature. 
The upper panels of Fig.~\ref{fig:dust} also show that the mid-IR 5--8$~\mu$m extinctions of amorphous olivine and pyroxene are considerably small compared to the 10$~\mu$m feature.
Hence, if a radiative transfer calculation is performed considering only silicate as the dust species, the wings of the silicate feature are likely to be seen in the emission while the peak in the absorption \citep{Kemper2002}. The typical extinction curves observed in our Galaxy \citep[e.g.,][]{Lutz1999, Indebetouw2005}, however, show relatively large mid-IR extinction as compared to the silicate features.
Therefore amorphous carbon \citep[ACH2;][]{Colangeli1995} is newly introduced in this study. The mid-IR absorption coefficient of amorphous carbon is relatively flat as shown in Fig.~\ref{fig:dust}, and thus we can enhance the 5--8$~\mu$m extinction compared to the 10$~\mu$m silicate feature by increasing the relative abundance of amorphous carbon. %as the extinction curves observed in our Galaxy \citep[e.g.,][]{Lutz1999, Indebetouw2005} calls for it.

We calculated the mass absorption and scattering coefficients, $\kappa_{\rm abs}$ and $\kappa_{\rm sca}$, of all the dust species except crystalline olivine, using the Distribution of Hollow Spheres \citep[DHS;][]{Min2003, Min2005} for the classical MRN dust size distribution \citep{Mathis1977}. The MRN dust size distribution is described by a power-law function as, 
\begin{equation}
    \frac{dn}{da} \propto a^{-3.5} ~~~~ (a_{\rm min} \leq a \leq a_{\rm max}),
\end{equation}
 where $a$ is the dust size. The minimum value of the dust size, $a_{\rm min}$, was fixed at 0.005$~\mu$m, while we tested two maximum values, $a_{\rm max}$, of 0.25$~\mu$m and 5.0$~\mu$m. 
% When $a_{\rm max}$ is 0.25$~\mu$m, $\kappa_{\rm abs}$ is much larger than $\kappa_{\rm sca}$ in the mid-IR wavelength range, and their spectral profiles are almost independent of the dust size. 
 %This is because 
 The former dust size distribution satisfies the Rayleigh limit in the mid-IR range. 
 \citet{Kemper2004} and \citet{Min2007} reproduced the silicate feature observed in the diffuse ISM in our Galaxy with the Rayleigh limit.
 DHS enables us to consider internal inhomogeneity by averaging the optical properties calculated for various volume fractions of the internal vacuum of the hollow sphere, $f$, for a uniform distribution within $0~{\leq}~f~{\leq}~f_{\rm max}$. 
We applied 0 and 0.7 to $f_{\rm max}$, and thus tested the four models as summarized in Table~\ref{tab:model}. $f_{\rm max}=0$ is equivalent to a homogeneous spherical dust model. %\citet{Min2007} reported that the DHS with 0.7 for $f_{\rm max}$ reproduces the extinction curve of the diffuse ISM in our Galaxy. 
For the purpose of calculating $\kappa_{\rm abs}$ and $\kappa_{\rm sca}$ of amorphous carbon, $a_{\rm max}$ and $f_{\rm max}$ were fixed at 0.25$~\mu$m and 0.7, respectively, which are expected for diffuse ISM silicate in our Galaxy, as the detailed properties of carbonaceous dust are out of scope in this study.

 The opacity of crystalline silicate shows sharp spectral features in the mid-IR range, and hence is much more sensitive to the grain morphology as well as the mineralogical composition than that of amorphous silicate.
% it is relatively easy to determine the grain size and the porosity as well as the mineralogical composition of crystalline silicate by analyzing the spectral profiles compared to amorphous silicate.
For example, it is known that a simple Mie calculation with a homogeneous spherical grain does not reproduce the astronomical or experimental data of crystalline silicate at all \citep[e.g., ][]{Fabian2001}. 
%, which suggests that the opacity of crystalline silicate is much more sensitive to the grain morphology than that of amorphous silicate. 
%\citet{Tsuchikawa2021} actually fit to the spectra of heavily obscured AGNs with the $\kappa_{\rm abs}$ of crystalline forsterite \citep[$\rm Mg_2SiO_4$; ][]{Tamanai2006}, which is the magnesium end member of the crystalline olivine group, reproduces the 11$~\mu$m crystalline feature of the PAH-less spectra well.
The Distribution of Form Factor (DFF) model, which is developed by \citet{Min2006}, is more flexible for the grain morphology than a similar statistical approach of the DHS. 
%より精度よくfitするために、我々はDFFを用いた。
In the DFF model, the mass absorption coefficient is calculated assuming the Rayleigh limit, as follows: 
\begin{equation}
{\kappa}_{\rm abs} = \frac{2\pi}{\rho\lambda}\int^1_0{\rm Im}\frac{P(L)}{1/(\epsilon-1)+L}dL,
\label{eq:dff}
\end{equation}
where $\rho$, $L$ and $P(L)$ are the mass density, the form factor and the form factor distribution, respectively. 
\citet{Zeidler2015} obtained $P(L)$ by fitting to the IR spectra of an olivine powder measured by \citet{Tamanai2006} with the optical constants of crystalline olivine (the San Carlos olivine; $\rm Mg_{1.72}Fe_{0.21}SiO_4$) using the DFF model. 
In the present study, we obtained the optical properties of the crystalline olivine with the DFF model assuming the optical constants measured at 300~K and $P(L)$ by \citet{Zeidler2015} for the three crystallographic axes, and averaged the optical properties over the axes.
%Here, we do not take into account variations in the crystalline features among the sample galaxies, whereas we partially investigate them in Appendix~\ref{app:lam23}.
The $\kappa_{\rm abs}$ is shown in the lower middle panel of Fig.~\ref{fig:dust}. We also obtained $\kappa_{\rm abs}$ and $\kappa_{\rm sca}$ of crystalline olivine for the wavelength range shorter than 7.5$~\mu$m in the same way as amorphous silicate with $a_{\rm max} = 0.25~{\rm {\mu}m}$, $f_{\rm max} = 0.7$ and the optical constants measured by \citet{Pitman2013}. 
For the purpose of the radiative transfer calculation, we gridded the three dust mass ratios of amorphous pyroxene to total amorphous silicate, $r_{\rm pyr}$, crystalline to total silicate, $r_{\rm cry}$, and silicate to total dust, $r_{\rm si}$. 
We calculated the dust continuum spectra, $F_{\rm DUSTY}$, for 151200 grid points for the geometrical and mineralogical parameter space of $p$, $Y$, $\tau_{\rm V}$, $r_{\rm pyr}$, $r_{\rm cry}$ and $r_{\rm si}$ as shown in Table~\ref{tab:param}.

\begin{deluxetable*}{lcccccc}
\tablenum{3}
\tablecaption{Dust properties used in this study}
\tabletypesize{\scriptsize} 
\label{tab:dust}
\centering  
\tablehead{
\colhead{Dust Species} & \colhead{Chemical Fomula} & 
\colhead{Mass Density} & \colhead{$a_{\rm max}$\tablenotemark{\tiny a}} & 
\colhead{$f_{\rm max}$\tablenotemark{\tiny b}} & \colhead{Wavelength Range} &
\colhead{$\rm Ref.$\tablenotemark{\tiny c}} \\ 
\colhead{} & \colhead{} & \colhead{(g/cc)} & \colhead{($\mu$m)} & 
\colhead{} & \colhead{(${\rm {\mu}m}$)} & \colhead{}
} 
\startdata
    \multirow{2}{*}{Amorphous Olivine} & 
    \multirow{2}{*}{$\rm MgFeSiO_4$} & 
    \multirow{2}{*}{3.71} & 
    \multirow{4}{*}{0.25~/~5.0} & 
    \multirow{4}{*}{0.0~/~0.7} & 0.01--0.20 & 1 \\ 
     & & & & & 0.20--500 & 2 \\
    \multirow{2}{*}{Amorphous Pyroxene} & 
    \multirow{2}{*}{$\rm Mg_{0.5}Fe_{0.5}SiO_3$} & \multirow{2}{*}{3.20} & & & 0.01--0.20 & 1\\
     & & & & & 0.20--500 & 2\\
    \tableline
    \multirow{2}{*}{Crystalline Olivine} & \multirow{2}{*}{$\rm Mg_{1.72}Fe_{0.21}SiO_4$}  &   \multirow{2}{*}{3.30} & 0.25 & 0.7 & 0.01--7.5 & 3\\ 
     & & & Rayleigh & -- & 7.5--500 & 4\\
    \tableline
    \multirow{2}{*}{Amorphous Carbon}  & 
    \multirow{2}{*}{C} & 
    \multirow{2}{*}{1.81} & 
    \multirow{2}{*}{0.25} & 
    \multirow{2}{*}{0.7} & 0.01--0.04 & 5\\
     & & & & &0.04--500 & 6\\
\enddata
\tablenotetext{\tiny a}{The maximum size of the MRN dust size distribution. The optical properties of crystalline olivine at the wavelengths longer than 7.5$~\mu$m are calculated with the Rayleigh limit.}
\tablenotetext{\tiny b}{The maximum vacuum fraction for DHS. The optical properties of crystalline olivine at the wavelengths longer than 7.5$~\mu$m are derived by a statistical method of DFF, not the DHS. }
\tablenotetext{\tiny c}{References for the optical constants: (1) \citet{Draine2021a}, (2) \citet{Dorschner1995}, (3) \citet{Pitman2013}, (4) \citet{Zeidler2015}, (5) \citet{Hagemann1975}, (6) \citet{Colangeli1995}}

\end{deluxetable*}

\subsection{Determination of the model parameters \label{sec:infer}}We estimated the model parameters which best reproduce the sample spectra, using the four models summarized in Table~\ref{tab:model} by the following steps: first, we conducted spectral fitting within narrow wavelength ranges to estimate the parameters of gas and ice absorption and PAH and line emission components. The maximum likelihood method with the Levenberg-Marquardt algorithm was used in the spectral fittings. In the subsequent steps, the amplitudes of the CO gas absorption and all the line emission components were fixed at the values obtained in the first step. Second, we performed the full-range spectral fitting of all the 151200 grid points in the geometrical and mineralogical parameter space shown in Table~\ref{tab:param}, and searched for the best-fit grid point. The same fitting algorithm as in the previous step was used. The ten parameters of the AGN-heated dust and the SF-heated dust and PAH emission components shown in Table~\ref{tab:param} were set to be free. We adopted the ice and PAH parameters obtained in the previous step as the initial parameters.

In the second step, the sampling of the parameter grid for the sake of the radiative transfer calculation is too coarse to estimate the uncertainties of the parameters from the probability distribution weighted by ${\rm exp}(-{\chi^2}/2)$ on the basis of the Bayesian statistics \citep[e.g., ][]{Kauffmann2003b}. Thus we performed a multi-dimensional log-linear interpolation of the parameter grid to approximate $F_{\rm DUSTY}$ in Eq.~\ref{eq:fagn} in a continuous parameter space, and determined the posterior probability distributions of all the parameters in Table~\ref{tab:param} using a Markov Chain Monte Carlo (MCMC) method. Emcee v3, which is a Python package implementing the affine invariant ensemble sampler \citep{Goodman2010, Foreman2013}, was used for the MCMC code. We applied uniform prior distributions for the plausible parameter ranges shown in Table~\ref{tab:param}. The MCMC was started from a small ball around the best-fit parameters obtained in the second step. We set an ensemble of 64 ``walkers'', which are the Markov chains evolving in parallel, and run the MCMC algorithm with 11000 steps, discarding the first 1000 steps as a ``burn-in'' period.
We adopted the best-fit step out of the 10000 steps for the resultant parameter set, which are used in the following sections. The 16th and 84th percentiles of the posterior distributions were adopted for the lower and upper uncertainties, respectively.

\begin{table*}[]
\tablenum{4}
	\caption{Summary of the four dust models used in the full-range spectral analysis}
	\label{tab:model}
    \centering       
	\begin{tabular}{llcc}
		\hline
		\hline
        Model & Description                     & $a_{\rm max}$ [$\mu$m] & $f_{\rm max}$ \\ 
		\hline
	    1 & homogeneous spherical/small-sized grain  & 0.25 & 0.0  \\
		2 & homogeneous spherical/large-sized grain   & 5.0  & 0.0  \\
		3 & porous/small-sized grain   & 0.25 & 0.7  \\
		4 & porous/large-sized grain    & 5.0  & 0.7  \\ 
		\hline
	\end{tabular}
\end{table*}

\section{Results\label{sec:result}}
\begin{figure*}
        \centering
        \includegraphics[width=14cm,clip]{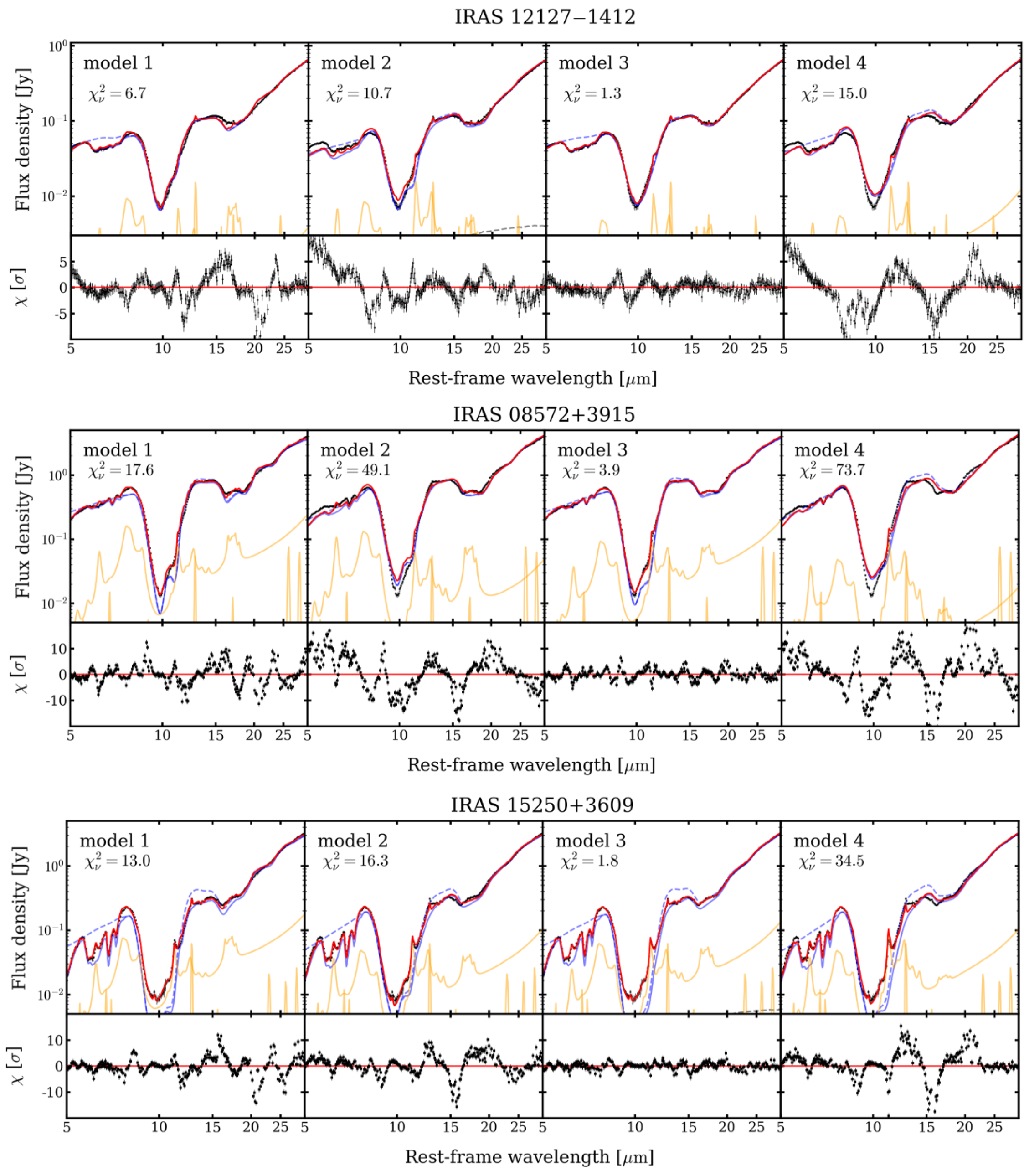}
        \caption{Examples of mid-IR spectral fits with models 1 (left), 2 (middle left), 3 (middle right) and 4 (right) to example spectra of IRAS~12127-1412, IRAS~08572+3915 and IRAS15250+3609. The best-fit spectrum by each model is shown with the red solid line. The blue dashed and solid lines represent the AGN-heated dust emission components without ice absorption, $F_{\rm d}^{\rm agn}$, and with ice absorption, $F_{\rm d}^{\rm agn} {\rm exp}(-\tau_{\rm ice})$, respectively, while the yellow lines represent the SF-heated dust and PAH emission, $F_{\rm sf}$, and the line emission components. %The gray dashed line shows the emission component due to the unobscured hot dust heated by AGN. 
        We show the reduced $\chi^2$ values in the upper left corner in each panel.}
        \label{fig:fit}%
   \end{figure*}

\begin{figure*}
        \centering
        \includegraphics[width=14cm,clip]{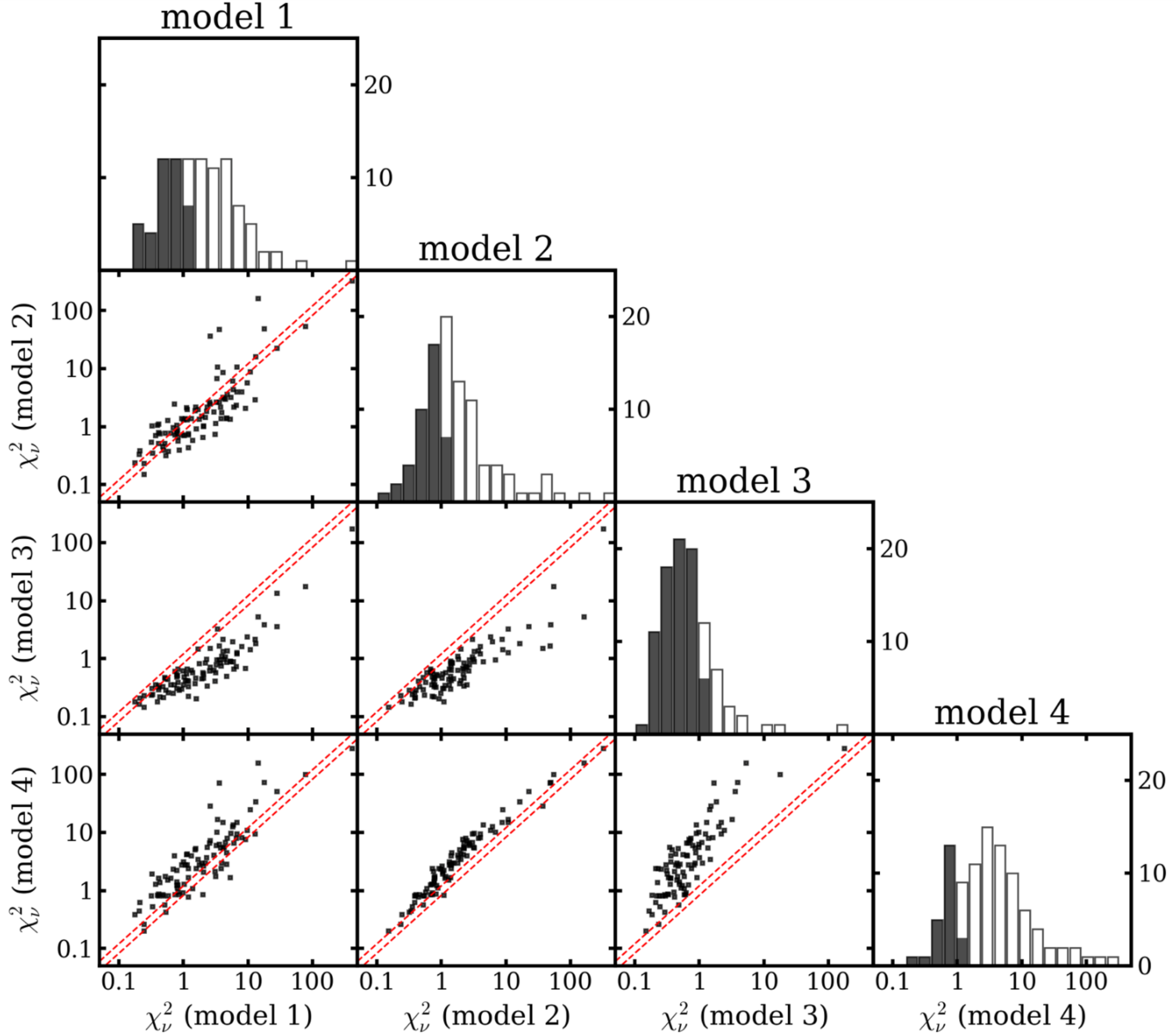}
        \caption{Comparison between the results of the four models. The panels at the diagonal positions show the histograms of reduced $\chi^2$ of each model. Black and white bars represent the objects, the fits of which are accepted and rejected, respectively, with a significance level of 5\%. The scatter plots at lower diagonal positions are the relationship between the reduced $\chi^2$ values of different two models. Red dashed lines show the thresholds that the $\chi^2$ between two models are significantly different with a significance level of 5\% on the basis of F-test.}
        \label{fig:chisq}%
   \end{figure*}

%\subsection{Properties of silicate dust obtained from the Spitzer/IRS full-spectral modeling}
We fitted all the spectra in our sample within 5--30$~\mu$m range with the four models, considering the dust radiative transfer effects. Figure~\ref{fig:fit} shows typical examples of the fitting results with the four dust models, indicating that all the four models well reproduce the overall profiles, such as the band ratio of the 10$~\mu$m to 18$~\mu$m features and the continuum shape, while the detailed quality of the fits is different between the models. 
We find that model~1 with $f_{\rm max}=0.0$ does not reproduce the central wavelength of the 18$~\mu$m amorphous feature, while model~3 with $f_{\rm max}=0.7$ reproduces it well owing to the peak shift of the 18$~\mu$m feature. % due to the porous structure. 
This trend is commonly seen for most of our sample spectra. We compared the reduced $\chi^2$ values between the models in Fig.~\ref{fig:chisq}, which indicates that model~3 gives better fitting results than model~1 for 96 out of the 98 objects in our sample, and their differences are significant for 88 out of the 96 objects with a significance level of 5\% by F-test. 
%殆どの天体で単純なMie計算によるモデルではうまくあわず、

Alternatively, model~2, in which we consider larger dust sizes, also shows the peak shift of the 10 and 18$~\mu$m features as compared to that of model~1 in Fig.~\ref{fig:dust}. %Therefore, similarly to model~3, model~2 is fitted better than model~1 (see Fig.~\ref{fig:chisq}). 
However the fits by model~2 do not improve so much as those by model~3. Comparing the reduced $\chi^2$ values between all the spectra, 97 out of the 98 objects prefer model~3, the porous and small-sized dust model, to model~2, and their differences are significant for 88 out of the 97 objects. This is likely owing to high scattering efficiency of large-sized dust as shown in Fig.~\ref{fig:dust} as well as a peak shift smaller than model~3. In the wavelength range where scattering is more effective, a photon travels a longer distance in the cloud on average before escaping the cloud, and thus the dust extinction is enhanced. Indeed, Fig.~\ref{fig:fit} shows that the best-fit ones of models~2 and 4, in which larger dust sizes are considered, deviate from the observed spectra at around 5 and 12$~\mu$m where the scattering coefficients are relatively high. %Though the fit to IRAS~15250+3609 by model~2 does not show such a deviation, either, model~3 is more plausible because the fit by model~2 can hardly reproduce the 12~$\mu$m absorption due to the $\rm H_2O$ ice libration mode in spite of the large ice absorption at 6$~\mu$m. %Note that larger-sized dust contributes to forward scattering more, although DUSTY assumes the isotropic dust scattering. The scattering asymmetry factor $g$ of amorphous olivine in models~2 and 4 is $\sim$0.5 at the wavelength of 10$~\mu$m. Therefore the forward scattering reduces the dust extinction, but is not likely to eliminate the deviation at 12$~\mu$m completely. 

\begin{figure}
    \figurenum{5}
    \plotone{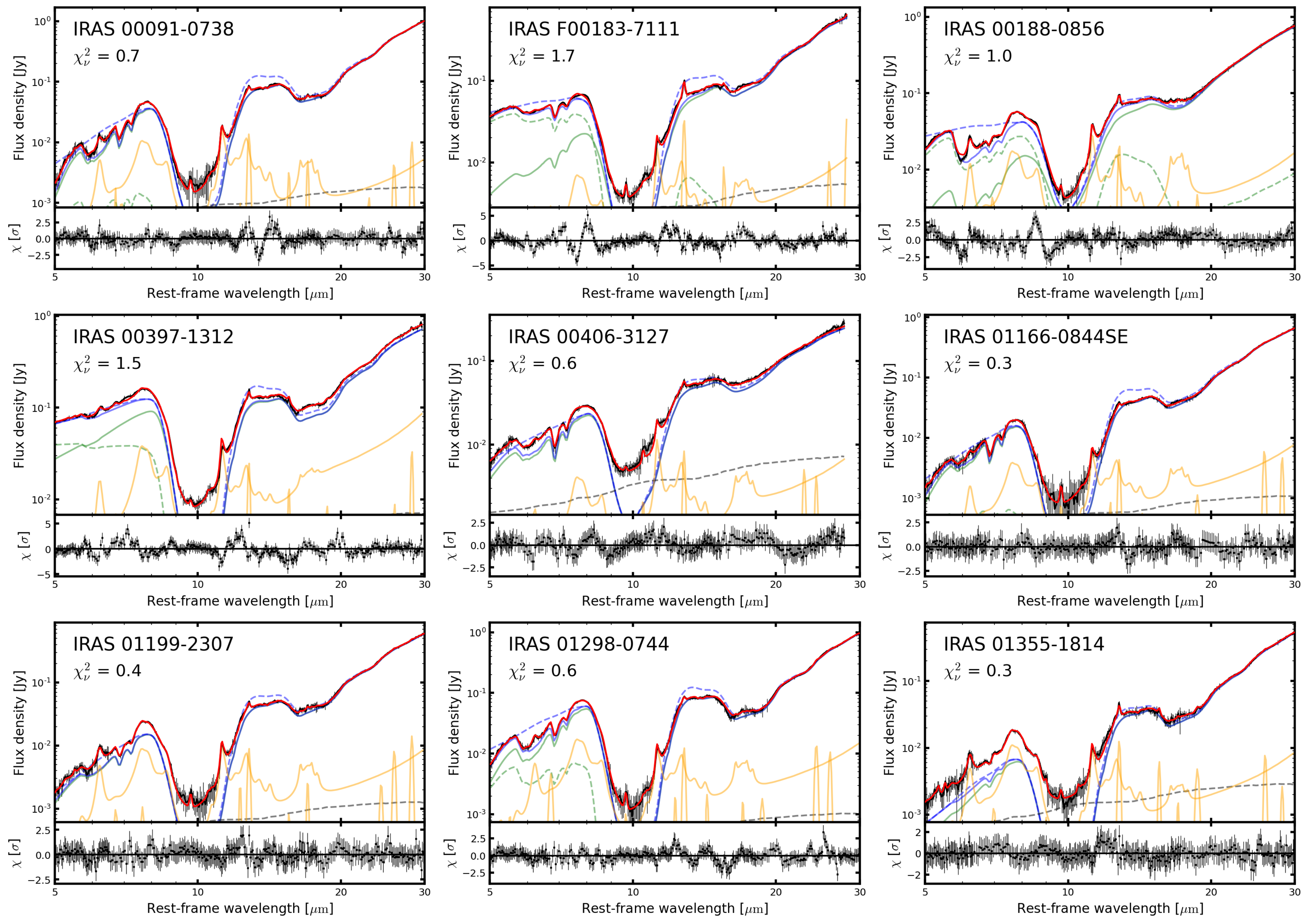}
    \caption{Examples of the modeling results of the mid-IR 5--30$~\mu$m spectra of heavily obscured AGNs. The best-fit spectrum is shown with the red solid line. The blue dashed and solid lines represent the AGN-heated dust emission components without ice absorption, $F_{\rm d}^{\rm agn}$, and with ice absorption, $F_{\rm d}^{\rm agn} {\rm exp}(-\tau_{\rm ice})$, respectively, while the yellow lines represent the SF-heated dust and PAH emission, $F_{\rm sf}$, and the line emission components. The AGN-heated dust emission component (blue solid line) is decomposed to the two components, the dust emission calculated by DUSTY (green solid line) and the additional obscured hot dust emission (green dashed line). The gray dashed line shows the emission component due to the unobscured hot dust heated by AGN.
    We show the reduced $\chi^2$ values in the upper left corner in each panel. The complete figure set (98 images) is available in the online journal.}
    \label{fig:fit_example}%
\end{figure}

We also confirm that model~4 cannot reproduce the sample spectra better than model~3.
In conclusion, model~3, which assumes the small size and porous structure, best reproduces 95 out of the 98 spectra of heavily obscured AGNs in our sample. Although models~1 and 2 are preferred by 2 and 1 out of the 98 objects, respectively, the preference is not statistically significant. 
Figure set~\ref{fig:fit_example} shows the modeling results of the full-range Spitzer/IRS spectra in our sample with model~3. %, and all the results are shown in Appendix. 
We can recognize that model~3 reproduces our sample spectra considerably well. 
Hence we use the results of model~3 for all the objects in the sample below.

  In addition to the dust size and the porosity, we obtain the mineralogical composition of amorphous silicate and the crystallinity with model~3. 
The pyroxene mass fraction, $r_{\rm pyr}$, and the crystallinity, $r_{\rm cry}$, thus obtained are summarized in Appendix. %, and all the fits by model~3 are shown in Fig.~\ref{fig:all}. 
Figure~\ref{fig:hist} shows the histograms of the dust properties of the $r_{\rm pyr}$ and $r_{\rm cry}$ in this study. 
The histogram of $r_{\rm pyr}$ shows a distribution skewed to the left, the 16th, 50th (median) and 84th percentiles of which are 0.5, 5.1 and 11.6\%, respectively.
The histogram of $r_{\rm cry}$ shows a relatively uniform distribution in a range of 0--14\%, the 16th, 50th (median) and 84th percentiles of which are 3.0, 5.8 and 8.2\%, respectively. 
\citet{Tsuchikawa2021} also derived the abundances of the amorphous pyroxene and crystalline silicate, in which the radiative transfer effects were not considered, but a simple full screen obscuration by dust was assumed despite the overprediction of the apparent optical depth of the 18$~\mu$m silicate feature compared to the 10$~\mu$m feature.
In addition, small distortions of the feature profiles due to the radiative transfer effects are not considered in the previous study at all. 
\citet{Tsuchikawa2021} obtained the crystallinity, by comparing the 10$~\mu$m amorphous and the 23$~\mu$m crystalline features. 
In principle, the radiative transfer affects the apparent optical depth ratio of two features with different wavelengths, as mentioned in Sect.~\ref{sec:intro} and \ref{sec:model}; however the difference in the degree of the radiative transfer effects between the sample galaxies is not taken into account in the previous results of crystallinity.
%Hence the previous results of crystallinity not considering the variation in the radiative transfer effect among the sample. 
In contrast, our analysis takes the radiative transfer into account, and therefore can reproduce the full-range IRS spectra.

\setcounter{figure}{5}
   \begin{figure*}
        \centering
        \includegraphics[width=14cm,clip]{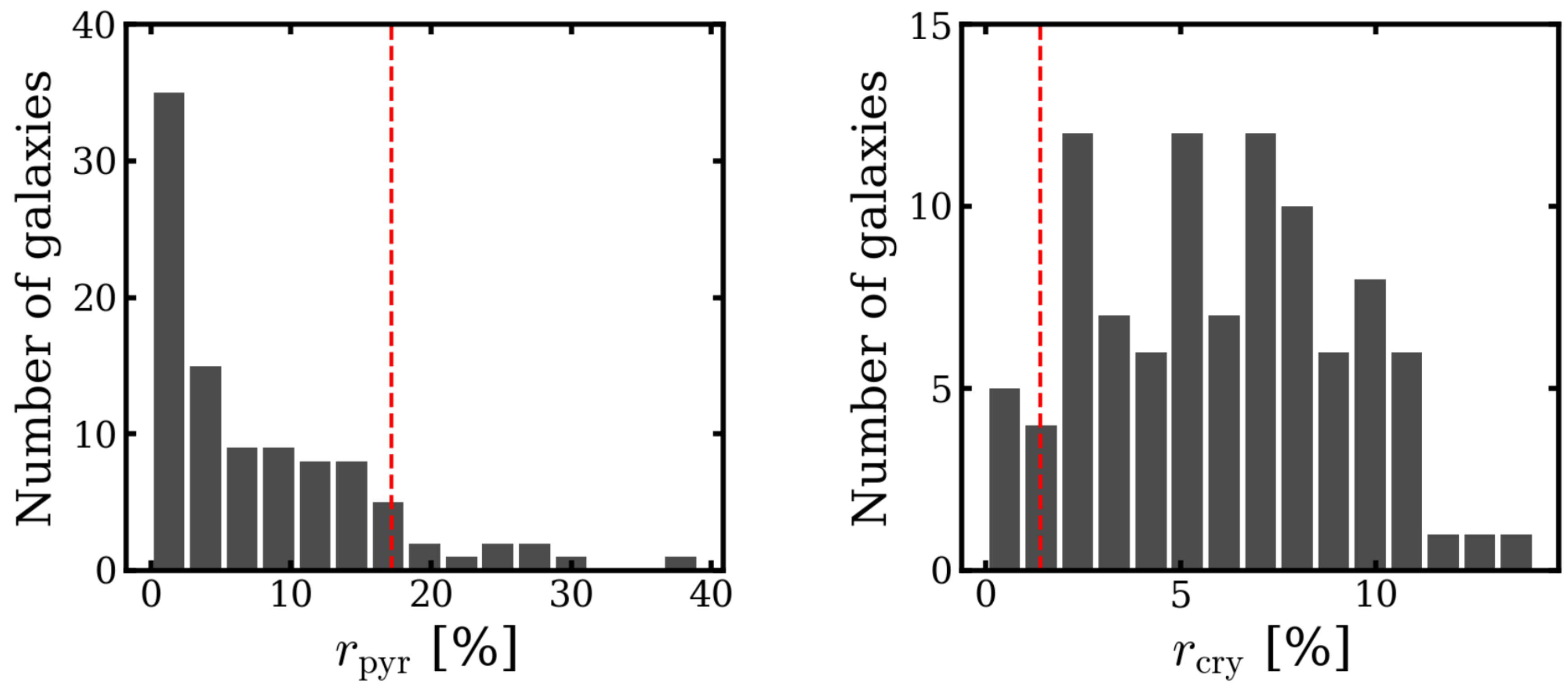}
        \caption{Histograms of the pyroxene mass fraction, $r_{\rm pyr}$, and the crystallinity, $r_{\rm cry}$. Red dashed lines correspond to typical dust properties of the diffuse ISM observed in our Galaxy \citep{DoDuy2020}. }
        \label{fig:hist}%
   \end{figure*}

\begin{figure*}
        \centering
        \includegraphics[width=14cm,clip]{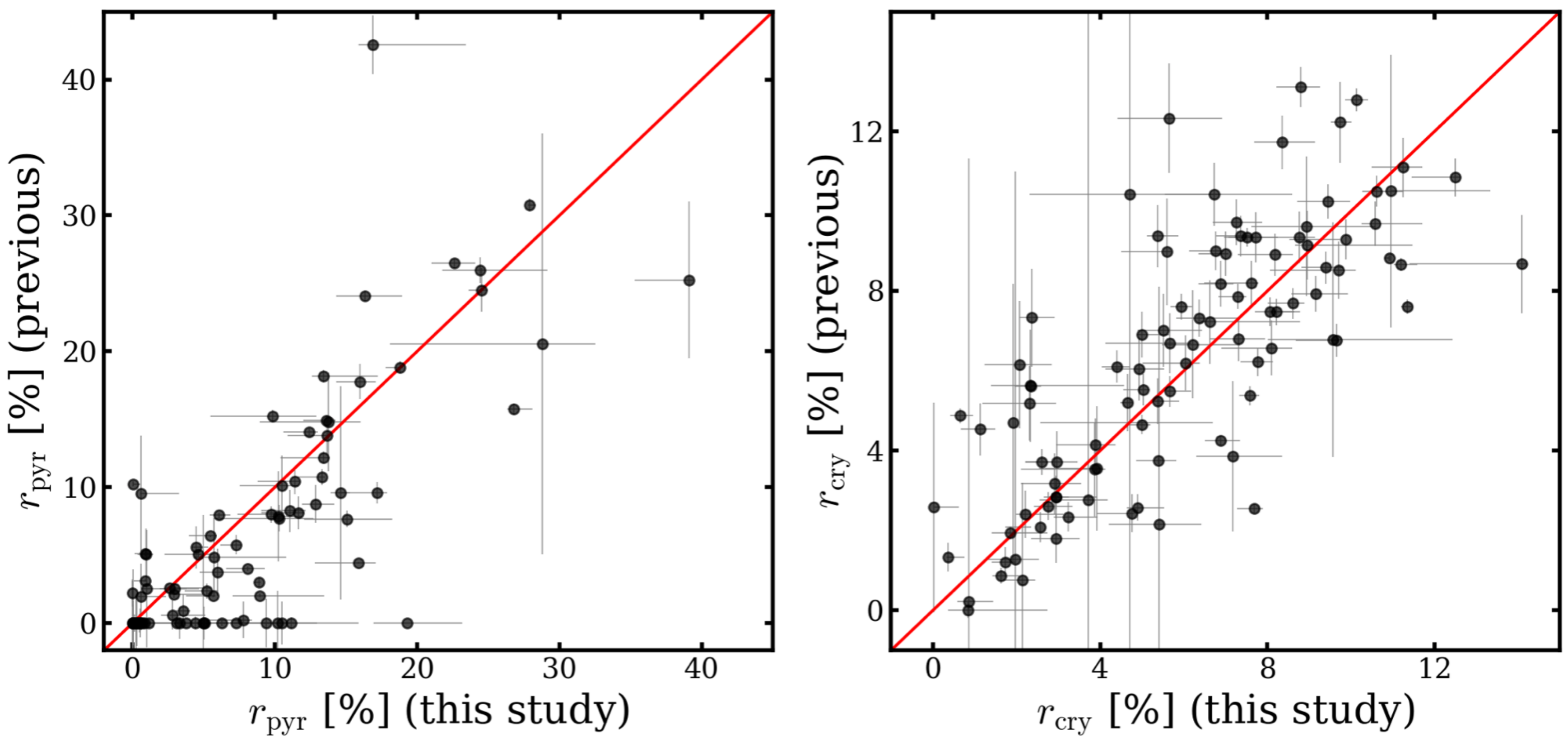}
        \caption{Correlation plots of the pyroxene mass fraction, $r_{\rm pyr}$, and the crystallinity, $r_{\rm cry}$, between the results of the previous and this studies. Note that \citet{Tsuchikawa2021} defined the pyroxene mass fraction with the total mass of amorphous and crystalline silicate, while the present study defines $r_{\rm pyr}$ with the total mass of amorphous silicate. Therefore we converted the pyroxene mass fraction obtained in \citet{Tsuchikawa2021} to the present definition in this figure.}
        \label{fig:vs2nd}%
   \end{figure*}

In order to verify the validity of the results by \citet{Tsuchikawa2021}, we compare both results in Fig.~\ref{fig:vs2nd}.
Note that the pyroxene mass fraction in the previous study is defined with the total mass of amorphous and crystalline silicate, while $r_{\rm pyr}$ in the present study is defined with the total mass of amorphous silicate. Therefore the pyroxene mass fraction is recalculated with the mass column densities of amorphous olivine and pyroxene obtained in the previous study according to the definition in the present study.
Figure~\ref{fig:vs2nd} shows that most of the objects are located around the diagonal lines. 
Thus the approximations by \citet{Tsuchikawa2021} are reasonable to some extent as a method to systematically analyze the silicate absorption feature of heavily obscured AGNs. Nevertheless, Fig.~\ref{fig:vs2nd} shows significant scatters from the diagonal lines. 
The scatters indicate that the degree of the radiative transfer effects, which reflect the geometrical distribution of the dust density and temperature, is different between the individual objects. Hence the present study is likely to derive the dust properties more reliably than the previous one.

\section{Discussion\label{sec:discuss}} 

We discuss overall properties of silicate dust in heavily obscured AGNs below. Variations of the properties among heavily obscured AGNs are shown and possible causes of the variations are mentioned in Appendix, where we cannot obtain significant ($r>0.7$; $r$ is the Pearson's correlation coefficient) relationships with environmental parameters.

\subsection{Amorphous silicate \label{sec:dis_amsil}}
The pyroxene mass fraction was obtained reliably by the full-range spectral modeling of the mid-IR spectra considering the radiative transfer effects. 
The red dashed lines in Fig.~\ref{fig:hist} correspond to the dust properties of the typical diffuse ISM in our Galaxy obtained in the sightline toward Sgr A* by \citet{DoDuy2020}.
It is found that 90\% of the sample galaxies show $r_{\rm pyr}$ lower than 17\% which is a value typical of the diffuse ISM in our Galaxy \citep{DoDuy2020}.
Thus amorphous silicate dust in heavily obscured AGNs tends to have notably pyroxene-poor mineralogy compared to that in the diffuse ISM in our Galaxy.

It is known that amorphous pyroxene tends to be extremely poor in the circumstellar environments of evolved stars \citep[e.g., ][]{Demyk2000, DoDuy2020}. The famous spectral profile of the silicate feature in the sightline toward the red supergiant $\eta$ Cep star is reproduced well by amorphous olivine alone \citep{Sargent2006}. \citet{Ferrarotti2001} theoretically predict the olivine-rich trend in the stellar ejecta of mass-loss stars with high mass-loss rates; olivine condensates earlier than pyroxene, and the radiation pressure on the olivine accelerates the wind material before pyroxene sufficiently condensates. %with the non-equilibrium dust condensation in the stellar winds. 
On the other hand, amorphous pyroxene as well as amorphous olivine is needed to reproduce silicate features as seen in molecular clouds and the circumstellar space around YSOs \citep[e.g.,][]{Demyk1999, vanBreemen2011, DoDuy2020}. From such trends, it is believed that amorphous silicate evolves from olivine to pyroxene in the ISM \citep{Demyk2001}. The conversion of amorphous olivine to pyroxene occurs due to the selective sputtering of oxygen atoms, for example, by cosmic-ray bombardments \citep[e.g.,][]{Demyk2001, Carrez2002, Rietmeijer2009}. Therefore silicate dust in heavily obscured AGNs on average is considered to be newly formed in the stellar ejecta and not processed much. On the basis of a merger-induced evolutionary scenario, obscuring materials of heavily obscured AGNs are likely to have been newly supplied by the recent nuclear starburst activities.

In addition, we investigate two kinds of the properties of amorphous silicate, which are the size distribution and the porosity.
It is concluded that the porosity of silicate dust needs to be introduced since the DHS model ($f_{\rm max}=0.7$) is favored as a result of the full-range spectral modeling. 
For various astronomical objects in our Galaxy, such as the diffuse ISM, YSOs, evolved stars and comets, the peak wavelength of the amorphous 18$~\mu$m feature cannot be reproduced well by a homogeneous spherical dust model calculated by the simple Mie theory but the DHS model or other models taking into account an internal inhomogeneity \citep[e.g.,][]{Min2007}. Hence silicate dust in heavily obscured AGNs is probably of high porosity as well.

It is also found that the sample spectra of heavily obscured AGNs do not suggest micron-sized large silicate dust. Large-sized silicate is often observed in circumstellar disks on the basis of the peak shifts of the mid-IR silicate features \citep[e.g.,][]{vanBoekel2005, Juhasz2010}. %, which is attributed to grain growth in the high-density gas environments. 
Large-sized silicate dust is also introduced to reproduce the mid-IR spectra of optically classified type-1 AGNs or QSOs \citep[e.g., ][]{Smith2010, Shi2014, Xie2017}, while it is not likely to be needed for optically classified type-2 AGNs because the peak wavelengths of their silicate features tend not to shift \citep[e.g.,][]{Hatziminaoglou2015}. %which can be also explained by the grain growth in the dense environments. %Alternatively, silicate is possibly processed in the intense radiation field.
The difference in the grain size estimated from the mid-IR silicate features is likely to depend on regions where the features originate. Actually, most of silicate features suggesting grain growth in optically classified type-1 AGNs, QSOs and circumstellar disks are observed as emission, not absorption, which are originated from dust located in inner hot regions. 
Hence the silicate absorption features of optically classified type-2 AGNs and heavily obscured AGNs reflect properties of dust located in relatively outer cool regions, where grain growth is not likely activated because gas density is lower than that in inner hot regions.
Accordingly, the small-sized and porous silicate which is suggested by the full-range modeling consistently supports the scenario that the pyroxene-poor silicate which obscures the AGNs is newly formed, originating from mass-loss stars in the recent starburst activity.

%The peak shift of the 10$~\mu$m silicate emission feature is also observed in the type-1 AGNs or QSOs \citep[e.g., ][]{Shi2014,Hatziminaoglou2015,Xie2017}. The UV and optical observations of QSOs indicate gray extinction curves \citep{Gaskell2004}. These observations suggest the existence of large-sized dust around AGNs, which are considered to be caused by, for example, selective destruction of the small grain \citep{Tazaki2020}. Therefore the silicate absorption observed in heavily obscured AGNs is likely due to relatively new dust which is not processed in such high density or harsh environments, from which we speculate that heavily obscured AGNs are in early evolutionary stages of AGNs relative to the type-1 AGNs or QSOs. 

\subsection{Crystalline silicate\label{sec:dis_crysil}}

In our Galaxy, crystalline silicate is abundant only in the circumstellar space around evolved stars and YSOs, the crystallinity of which are 10--15\% and a few--40\%, respectively \citep{Henning2010}; the ISM silicate in our Galaxy is known to show almost no crystalline signature. For example, \citet{DoDuy2020} and \citet{Min2007} obtained low mass abundances of crystalline silicate, $1.4\pm0.2\%$ and $\sim1\%$, respectively, in the line of sight toward Sgr A*. 
Therefore the cosmic silicate is considered to be rich in the crystalline phase only immediately after the mass ejection by mass-loss stars, and then rapidly processed to the amorphous phase in the interstellar spaces due to cosmic-ray bombardments \citep[e.g.,][]{Demyk2001,Kemper2004,Bringa2007}. In the circumstellar space of YSOs such as protoplanetary disks, amorphous silicate is supplied from the interstellar space, and then crystallized due to thermal annealing in the hot environment close to the central star \citep[e.g.,][]{Gail2004}.

We find that the crystallinity in heavily obscured AGNs distributes widely from 0\% to 14\%. 
Based on the scenario that silicate in heavily obscured AGNs is relatively fresh dust through the recent starburst activity, the crystallinity higher than 10\% in the sample is expected to be attributed to the silicate originating from mass-loss stars. However, cosmic ray bombardments can completely amorphize the crystalline silicate in a short timescale of $\sim$70~Myr in our Galaxy \citep{Bringa2007}. Indeed, \citet{Kemper2011} simulated whether or not such high crystallinity can be achieved in starburst galaxies considering the balance of production, destruction and amorphization of silicate. They concluded that the crystallinity higher than 10\% observed in heavily obscured AGNs is difficult to be reproduced by the starburst activity alone. Thus it is likely that the silicate dust in the heavily obscured AGNs is not only just newly formed through the starburst activity but also crystallized later by other mechanisms, such as thermal annealing.

Amorphous silicate crystallizes at around 1000~K. 
On the other hand, in heavily obscured AGNs, since the mid-IR 23~$\mu$m spectral bands due to crystalline silicate are detected only in the absorption, the crystalline silicate is likely to be located in outer cool regions compared to the vicinity of nucleus, as also pointed out by \citet{Spoon2006}. 
A candidate mechanism of re-crystallization in cooler regions is an in-situ crystallization due to transient heating caused by shock waves \citep{Harker2002}, which can be driven by outflows. Indeed, outflows are detected in many of the sample galaxies, and shock heating is predicted, for example, through mid-IR pure rotational molecular hydrogen lines \citep[e.g.,][]{Hill2014}.   
An annealing time of the transient heating is expected to be so short that the chemical equilibrium cannot be achieved, and hence crystalline enstatite, the Mg end member of crystalline pyroxene, is unlikely to be formed \citep{Gail2004}. Therefore no detection of crystalline enstatite, which peaks at the wavelengths of e.g., 18.5, 28 and 36 $\mu$m, in our sample spectra reasonably supports the in-situ crystallization.

For another re-crystallization mechanism of amorphous silicate in cooler regions in heavily obscured AGNs, we consider that amorphous silicate is crystallized in the high temperature environments in the vicinity of nucleus and then transported to an outer cooler region. 
Indeed, crystalline forsterite is detected in a quasar wind spectrum \citep{Kemper2007}.
However, \citet{Spoon2006} mentioned that the re-crystallization mechanism is not plausible because a large-scale transportation mechanism itself needs to be introduced. Nevertheless, as already mentioned above, recent
observations reveals that outflows are ubiquitously present in the nuclear regions of
ULIRGs \citep[e.g.,][]{Veilleux2020}.
Therefore circumnuclear material may have experienced a large-scale transportation,  and thus crystalline silicate is possibly transported. 
For the high crystallinity in spite of the large-scale transportation, the timescale of amorphization must be longer than that of the outflow transportation to the outer cool region.
For instance, the presence of a highly collimated molecular jet in the nuclear region of NGC~1377 is reported with ALMA \citep{Aalto2012, Aalto2016, Aalto2020}.
The scale and age of the outflow are found to be 200~pc and 1.4~Myr, respectively \citep{Aalto2012}.
%それに対して、The torus-like 0.8~mm dust continuum is observed at scale of 20~pc \citep{Aalto2016}. In addition, 
On the other hand, the dust temperature at inner 3~pc is likely to be $\sim$180~K, which should contribute to the mid-IR continuum emission. Comparing these outflow parameters with a typical amorphization timescale of 70~Myr due to cosmic-ray bombardments in our Galaxy \citep{Bringa2007}, the size of the mid-IR emitting layer is small enough to supply silicate dust with the high crystallinity by the transportation.

\section{Conclusions}
Various properties of silicate dust are imprinted on the profiles of the silicate absorption features observed in the mid-IR spectra of heavily obscured AGNs. 
We have selected 98 heavily obscured AGNs which show notably deep silicate absorption features in the mid-IR spectra observed by Spitzer/IRS. 
From the sample spectra, properties of the silicate dust in heavily obscured AGNs have been estimated systematically by the full-range 5--30$~\mu$m spectral modeling. 
The properties of silicate dust thus obtained are the pyroxene mass fraction $r_{\rm pyr}$, the crystallinity $r_{\rm cry}$, the size distribution and the porosity. 
%As an additional piece of information, we also focused on the peak wavelength of the 23$~\mu$m crystalline band $\lambda_{23}$.
The results obtained in this study are summarized below.

\begin{itemize}
      \item Three dust species of amorphous olivine, amorphous pyroxene and crystalline forsterite are needed to account for the differences in the silicate features. Their composition ratios widely vary among the sample galaxies. 
      The median $r_{\rm pyr}$ is 5.1\%, while several sources show significantly high values around 30\%.
      The $r_{\rm cry}$ almost uniformly distributes in a range of 0--14\%.

      \item Comparing the results of the mid-IR full-range spectral modelings between four dust models with different sizes and porosities, 97\% of the sample galaxies prefer the porous silicate dust model without micron-sized large grains. 
      $r_{\rm pyr}$ and $r_{\rm cry}$ obtained by the full-range spectral modeling are overall consistent with the results obtained by the narrow-range 5.3--12$~\mu$m spectral modeling, but more reliable than the latter results for the individual galaxies.  %On the other hand, the correlations between the dust properties obtained by the two modelings scatter largely, which is attributed to the difference in the accuracies of the parameter estimates with or without considering the radiative transfer effects.
      
    %  \item $\lambda_{23}$ varies in a range of 23 to 24$~\mu$m. In particular, NGC~1377 and IRAS~F13279+3401 show notably long $\lambda_{23}$ of 23.86$\pm$0.11 and 23.53$~\pm$0.04$~\mu$m, respectively, which can be caused by high porosity of crystalline silicate.

\end{itemize}

 Comparing the overall dust properties in heavily obscured AGNs with those in our Galaxy, we discuss the origin of the properties of silicate in heavily obscured AGNs, as summarized below.

\begin{itemize}

    \item The overall pyroxene-poor mineralogical composition, small dust size and porosity of silicate dust in heavily obscured AGNs are similar to the circumstellar silicate ejected from mass-loss stars in our Galaxy. This trend suggests that silicate in heavily obscured AGNs is newly formed dust, which is presumably due to the recent circumnuclear starburst activity considering the merger-induced evolutionary scenario.
      
    \item
    The crystalline silicate in heavily obscured AGNs is likely to be located in outer regions cooler than the crystallization temperature, since 23$~\mu$m crystalline band in heavily obscured AGNs is detected only in the absorption. In order to explain $r_{\rm cry}$ higher than 10\%, we propose two crystallization scenarios: one is the combination of thermal processing in the center and radial transportation of the crystalline silicate by outflows. The other is the in-situ transient heating of silicate by shocks originating from outflows.

\end{itemize}

\begin{acknowledgements}
This study is based on observations with the Spitzer Space Telescope, which is operated by the Jet Propulsion Laboratory, California Institute of Technology under a contract with NASA, using the Combined Atlas of Sources with Spitzer IRS Spectra (CASSIS). CASSIS is a product of the IRS instrument team, supported by NASA and JPL. This study was supported by Grant-in-Aid for JSPS Fellows No.~21J14438. 
\end{acknowledgements}

\vspace{5mm}
\facilities{Spitzer(IRS)}

%% Similar to \facility{}, there is the optional \software command to allow 
%% authors a place to specify which programs were used during the creation of 
%% the manuscript. Authors should list each code and include either a
%% citation or url to the code inside ()s when available.

\software{DUSTY \citep{Ivezic1997}, 
          Emcee v3\citep{Goodman2010, Foreman2013}
          }

\appendix
\section{Variations in the properties of silicate dust in heavily obscured AGNs\label{sec:variation}}

We find that there are significant variations in the pyroxene mass fraction, $r_{\rm pyr}$, and the crystallinity, $r_{\rm cry}$. %, and the peak position of the 23$~\mu$m feature, $\lambda_{23}$. 
The variations imply the processing of silicate dust in different environments among the individual heavily obscured AGNs. Therefore we compare them with the statistical properties of the sample galaxies to better comprehend the nature of the properties of silicate dust.

\subsection{Evolutionary scenarios \label{sec:dis_lir}}

\begin{figure*}
\centering
\includegraphics[width=16cm,clip]{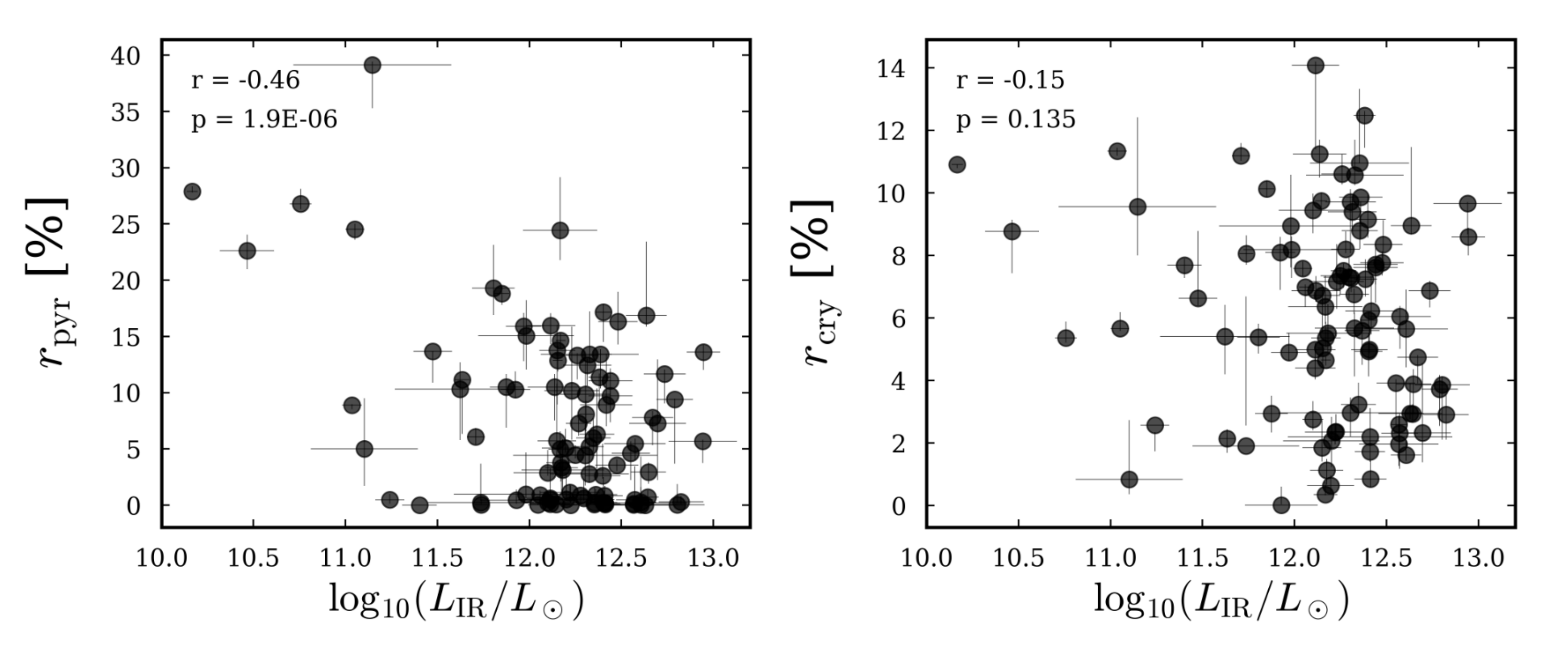}
\caption{Properties of silicate dust as a function of the total infrared luminosity, $L_{\rm IR}$, shown together with the Pearson's correlation coefficient and the p-value in the top-left corner.}
\label{fig:dis_lir}%
\end{figure*}

Table~\ref{tab:sample} summarizes the total 8--1000$~\mu$m IR luminosity, $L_{\rm IR}$, which generally reflects the total power of SF and AGN activities, for 97 out of all the 98 sample galaxies. 
$L_{\rm IR}$ in the sample galaxies ranges widely over three orders of magnitude from $1.5\times10^{10}~L_{\odot}$ for NGC~1377 to $8.8\times10^{12}~L_{\odot}$ for IRAS~F00183-7111. 
Figure~\ref{fig:dis_lir} shows the relations of the dust properties with $L_{\rm IR}$, indicating a negative tendency between $r_{\rm pyr}$ and $L_{\rm IR}$ ($r=-0.46$ and $p=1.9{\times}10^{-6}$, where $r$ and $p$ correspond to the Pearson's correlation coefficient and the $p$-value, respectively), while $r_{\rm cry}$ does not show any tendency with $L_{\rm IR}$ ($r=-0.15$, $p=0.14$). 
As shown in the left panel of Fig.~\ref{fig:dis_lir}, the negative tendency is likely to be attributed to the low-$L_{\rm IR}$ galaxies with ${\rm log}_{10}(L_{\rm IR}/L_{\odot}) < 11.2$. 
There are 6 pyroxene-rich sources which show $r_{\rm pyr}>20\%$ in Fig.~\ref{fig:dis_lir}; only one of them has relatively high $L_{\rm IR}$, while all the other sources have low $L_{\rm IR}$ (${\rm log}_{10}(L_{\rm IR}/L_{\odot}) < 11.2$).
Thus silicate dust in the low-$L_{\rm IR}$ obscured AGNs is expected to be relatively old according to the mineralogical evolutionary picture \citep[see Sect.~\ref{sec:dis_amsil};][]{Demyk2001}.
Relatively high-$L_{\rm IR}$ sources probably evolve through gas-rich mergers, whereas the low-$L_{\rm IR}$ sources are unlikely to experience such gas-rich mergers \citep{Blecha2018, Kim2021}. 
Taking into account that gas-rich mergers activate the circumnuclear star formations, low-$L_{\rm IR}$ obscured AGNs are likely to be deficient in newly formed dust, which can reasonably explain that the low-$L_{\rm IR}$ obscured AGNs have relatively old silicate dust.

\subsection{Evolutionary stages \label{sec:dis_tau}}

\begin{figure*}
\centering
\includegraphics[width=16cm,clip]{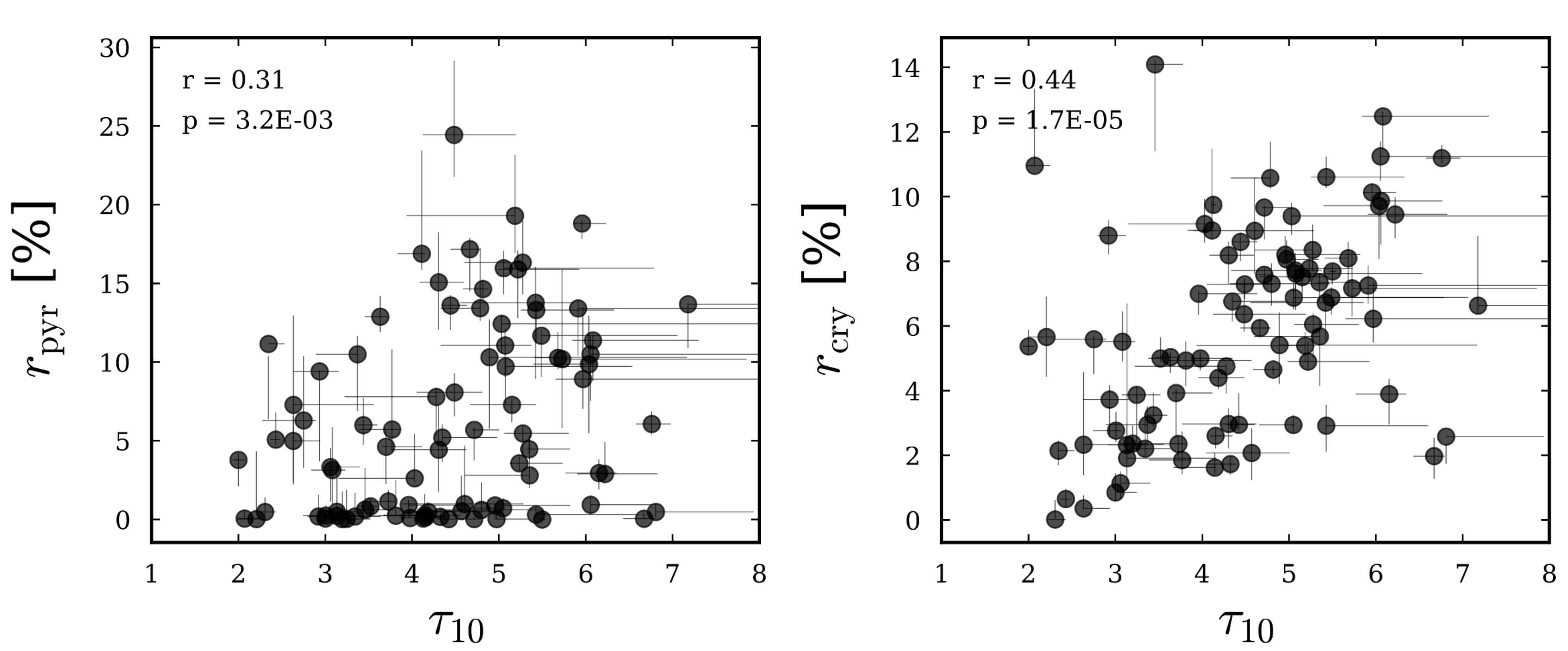}
\caption{Relations of the properties of silicate with the optical depth of the 10$~\mu$m silicate absorption feature, $\tau_{\rm 10}$, shown only for high-$L_{\rm IR}$ sources with ${\rm log}_{10}(L_{\rm IR}/L_{\odot}) > 11.2$. The Pearson's correlation coefficient and the p-value are shown in the top-left corner.}
\label{fig:dis_tau}%
\end{figure*}

As the evolutionary stage advances, ionized gas outflows should be developed, resulting in less obscuration.
Therefore the optical depth of the silicate absorption feature can be regarded as an indicator of the evolutionary stage of heavily obscured AGNs. 
Indeed, \citet{Spoon2009a} and \citet{Spoon2009b} found that the two galaxies, IRAS~F00183-7111 and IRAS~12127-1412, which are thought to be in relatively advanced evolutionary stages because of the signatures of the ionized gas outflows, show mild silicate absorption compared to other heavily obscured AGNs in the mid-IR classification diagram of \citet{Spoon2007}. %the galaxies in the diagonal branch, which to the . 
Figure~\ref{fig:dis_tau} shows the relation between the dust properties of the relatively high-$L_{\rm IR}$ sources with ${\rm log}_{10}(L_{\rm IR}/L_{\odot}) > 11.2$, which do not show unusually high $r_{\rm pyr}$ in Fig.~\ref{fig:dis_lir}, and the optical depth of the 10$~\mu$m silicate absorption feature, $\tau_{\rm 10}$, as summarized in Table~\ref{tab:result}.
$\tau_{\rm 10}$ is calculated in the same way as performed for the sample selection in the previous and present studies but using the best-fit model obtained in this study with the corrections for the PAH, line and unobscured emission and ice absorption components. 
It is found that $r_{\rm pyr}$ and $r_{\rm cry}$ show positive tendencies ($r=0.31$, $p=0.0032$ and $r=0.44$, $p=1.7{\times}10^{-5}$, respectively). 
Thus we consider a possible picture that the amorphous pyroxene and crystalline silicate are less abundant in later evolutionary stages for the relatively high-$L_{\rm IR}$ sources which were likely to have experienced the gas-rich merger and starburst activity.

The mineralogical evolutionary scenario predicts that pyroxene-richer silicate is relatively older \citep{Demyk2001}, as already mentioned in Sect.~\ref{sec:dis_amsil}. 
Therefore high-$L_{\rm IR}$ sources in later stages are possibly lacking in old amorphous silicate. 
This can be reasonably explained by a dynamical evolutionary picture that pyroxene-rich old silicate could be blown out by outflows with the evolutionary time, and replaced by the newly formed pyroxene-poor silicate which originates from the recent starburst activity and has been supplied into the obscuring clouds.

The trend that the crystallinity decreases as $\tau_{10}$ decreases can be explained by considering the X-ray-induced amorphization \citep{Ciaravella2016,Gavilan2016}. 
In our Galaxy, \citet{Glauser2009} reported negative correlations of the X-ray luminosity and hardness with the crystallinity of the circumstellar silicate associated with class~II YSOs.  
The amount of the material absorbing X-ray from the central engine decreases with the evolutionary time. In the case of heavily obscured AGNs, X-ray fluxes in the nuclear region can also increase with the SMBH growth. Accordingly, the X-ray photons can affect a larger fraction of the obscuring clouds in later evolutionary stage.

\subsection{Orientation effects \label{sec:dis_ori}}

A spherically symmetric distribution is assumed for the density and temperature of the circumnuclear dust in the spectral modeling in this study for simplicity. In reality, the existence of the outflows and the AGN unified scheme call for an axisymmetric but not spherically symmetric structure, as the crystallization scenarios proposed in Sect.~\ref{sec:dis_crysil} are associated with outflows.
Hence it is important to examine the dependence on the orientation to confirm the validity of the scenarios. 

\begin{figure*}
\centering
\includegraphics[width=16cm,clip]{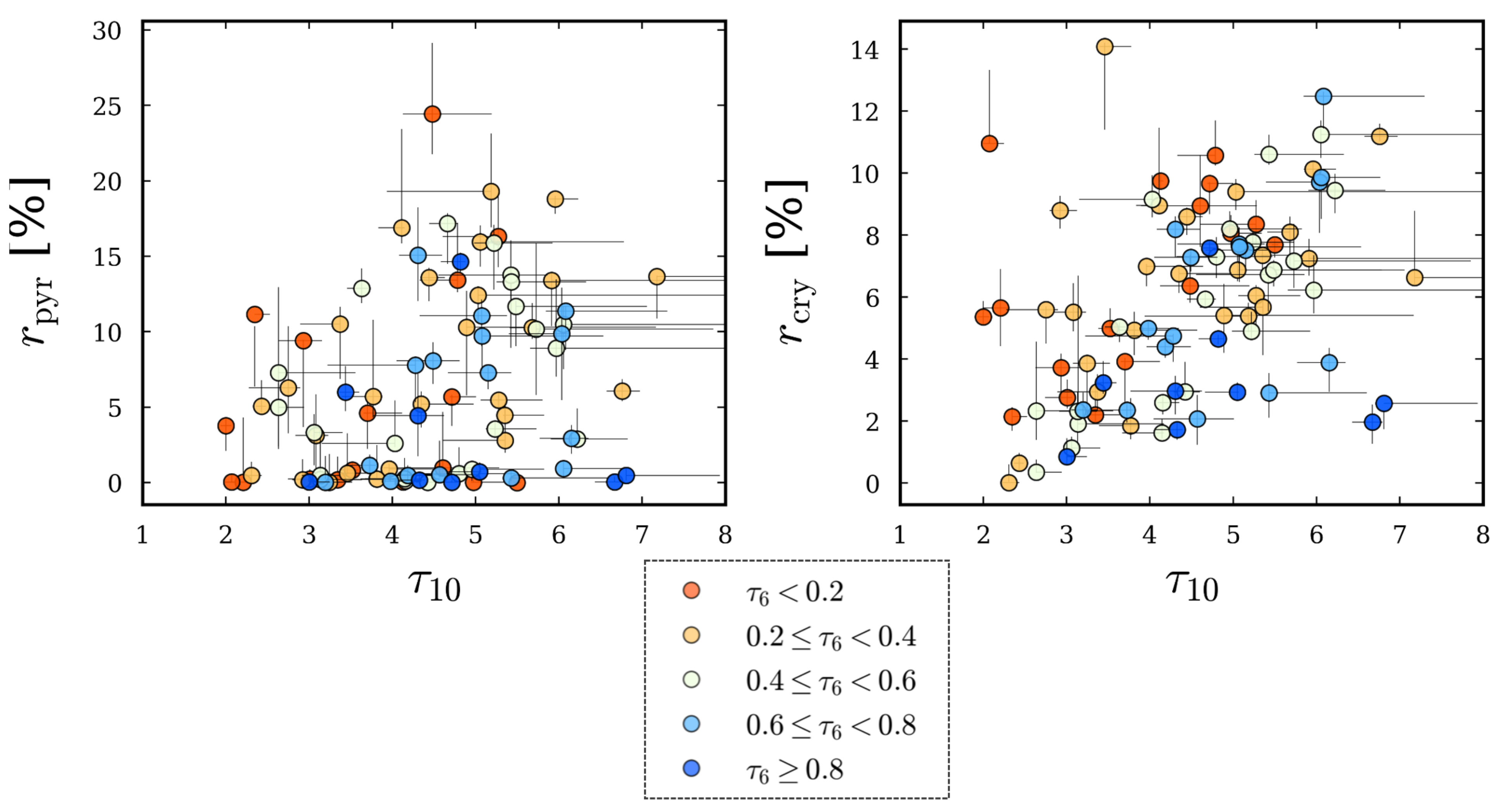}
\caption{Same as Fig.~\ref{fig:dis_tau}, but the data points are color-coded according to the optical depth of the 6$~\mu$m $\rm H_2O$ ice absorption, $\tau_{6}$. }
\label{fig:dis_ice}%
\end{figure*}

In Figure~\ref{fig:dis_ice}, we color-code the data points in Fig.~\ref{fig:dis_tau} according to the optical depth of the 6$~\mu$m $\rm H_2O$ ice absorption, $\tau_{6}$, as summarized in Table~\ref{tab:result}.
The figure exhibits systematic differences in the distribution of the data points between galaxies with small $\tau_{6}$ and those with large $\tau_{6}$. We perform a Kolmogorov-Smirnov test to verify the significance of the differences between the samples with $\tau_{6} < 0.4$ and $\tau_{6} > 0.6$. As a result, we find that the distribution of $r_{\rm cry}$ is different between the samples ($p = 0.023$), while that of $r_{\rm pyr}$ is not ($p = 0.18$).
$\rm H_2O$ ice can exist only in regions colder than the sublimation temperature of 90~K \citep{Tielens2005}. Such cold regions are likely to be located in the direction relatively perpendicular to the outflows since the collisional heating by shocks should be less efficient in a direction closer to the edge-on view. The X-ray heating is also inefficient in the edge-on direction because of the large gas column density. 
Therefore the trend of low $r_{\rm cry}$ for ice-rich sources suggests that crystalline silicate is richer in regions closer to the outflow directions, which can be consistently explained by the crystallization scenarios proposed in Sect.~\ref{sec:dis_crysil}.

\section{Summary of all the fitting results\label{app:summary}}

We perform the spectral modeling to determine the dust properties in heavily obscured AGNs. The pyroxene mass fraction, $r_{\rm pyr}$, and the crystallinity, $r_{\rm cry}$, thus obtained are summarized in Table~\ref{tab:result}, which also include the optical depths of the 10~$\mu$m silicate and the 6~$\mu$m $\rm H_2O$ ice absorption features $\tau_{10}$ and $\tau_{6}$, respectively. %All the fits of the full-range spectral modeling are shown in Fig.~\ref{fig:all}.

\startlongtable
\tablenum{5}
\begin{deluxetable*}{lcccc}
\tablecaption{Summary of the dust properties of the sample}
\tabletypesize{\scriptsize}
\label{tab:result}   
\centering    
\tablehead{
\colhead{Name} & \colhead{$r_{\rm pyr}$ [\%]} & 
\colhead{$r_{\rm cry}$ [\%]} & 
\colhead{$\tau_{10}$} & \colhead{$\tau_{6}$}
} 
\decimalcolnumbers
\startdata        IRAS~00091--0738 &   $7.28_{-1.10}^{+0.40}$ &   $7.52_{-0.54}^{+0.22}$ & $5.15_{-0.48}^{+0.28}$ & 0.69$~\pm~$0.04 \\
       IRAS~F00183--7111 &  $13.60_{-1.58}^{+0.65}$ &   $8.60_{-0.59}^{+0.29}$ & $4.44_{-0.10}^{+0.19}$ & 0.28$~\pm~$0.01 \\
        IRAS~00188--0856 &  $0.05_{-0}^{+0.50}$ &   $0.85_{-0.26}^{+0.59}$ & $3.00_{-0.00}^{+0.25}$ & 0.96$~\pm~$0.02 \\
        IRAS~00397--1312 &   $5.69_{-1.95}^{+0.13}$ &   $9.66_{-0.98}^{+0.05}$ & $4.72_{-0.02}^{+0.29}$ & 0.13$~\pm~$0.01 \\
        IRAS~00406--3127 &  $0.04_{-0}^{+1.85}$ &   $3.87_{-1.76}^{+0.26}$ & $3.25_{-0.17}^{+0.28}$ & 0.36$~\pm~$0.02 \\
      IRAS~01166--0844SE &  $15.96_{-1.66}^{+1.10}$ &   $6.88_{-0.36}^{+0.46}$ & $5.05_{-0.00}^{+1.73}$ & 0.38$~\pm~$0.04 \\
        IRAS~01199--2307 &  $12.43_{-1.84}^{+0.59}$ &   $9.40_{-0.59}^{+0.41}$ & $5.03_{-0.00}^{+3.14}$ & 0.34$~\pm~$0.04 \\
        IRAS~01298--0744 &   $0.93_{-0.58}^{+0.50}$ &  $9.87_{-1.34}^{+-0.13}$ & $6.06_{-0.00}^{+0.71}$ & 0.74$~\pm~$0.02 \\
        IRAS~01355--1814 &  $16.32_{-2.03}^{+2.64}$ &   $8.35_{-0.66}^{+0.79}$ & $5.27_{-0.67}^{+0.00}$ & 0.13$~\pm~$0.07 \\
        IRAS~F01478+1254 &   $0.98_{-0.34}^{+3.71}$ &   $8.94_{-1.32}^{+1.64}$ & $4.61_{-0.71}^{+0.68}$ & 0.12$~\pm~$0.09 \\
        IRAS~01569--2939 &  $13.30_{-2.12}^{+0.20}$ &  $10.61_{-0.34}^{+0.63}$ & $5.43_{-0.18}^{+0.90}$ & 0.40$~\pm~$0.03 \\
        IRAS~02455--2220 &   $7.29_{-5.07}^{+5.67}$ &   $2.33_{-0.94}^{+2.24}$ & $2.63_{-0.01}^{+0.93}$ & 0.46$~\pm~$0.17 \\
         IRAS~02530+0211 &  $24.52_{-0.92}^{+0.47}$ &   $5.66_{-0.07}^{+0.53}$ & $3.47_{-0.07}^{+0.13}$ & 0.00$~\pm~$0.00 \\
         IRAS~03158+4227 &  $0.27_{-0}^{+1.35}$ &   $1.62_{-0.19}^{+0.44}$ & $4.15_{-0.48}^{+0.02}$ & 0.54$~\pm~$0.04 \\
                NGC~1377 &  $27.88_{-0.10}^{+0.46}$ & $10.92_{-0.12}^{+-0.00}$ & $4.06_{-0.02}^{+0.02}$ & 0.02$~\pm~$0.01 \\
        IRAS~03538--6432 &   $9.40_{-5.72}^{+0.06}$ &   $3.72_{-1.17}^{+0.46}$ & $2.93_{-0.31}^{+0.22}$ & 0.19$~\pm~$0.03 \\
         IRAS~03582+6012 &  $0.00_{-0}^{+0.02}$ &   $7.69_{-0.41}^{+0.20}$ & $5.50_{-0.03}^{+0.04}$ & 0.15$~\pm~$0.00 \\
        IRAS~04074--2801 &   $4.46_{-0.30}^{+0.87}$ &   $7.35_{-0.58}^{+0.28}$ & $5.35_{-0.02}^{+0.47}$ & 0.36$~\pm~$0.02 \\
        IRAS~04313--1649 &   $7.78_{-2.46}^{+0.58}$ &   $4.75_{-0.84}^{+0.06}$ & $4.28_{-1.06}^{+0.00}$ & 0.66$~\pm~$0.10 \\
        IRAS~04384--4848 &  $0.23_{-0}^{+2.26}$ &   $4.93_{-0.80}^{+0.59}$ & $3.81_{-0.19}^{+0.76}$ & 0.37$~\pm~$0.05 \\
          ESO~203--IG001 &  $18.80_{-0.99}^{+0.20}$ &  $10.13_{-0.27}^{+0.27}$ & $5.96_{-0.11}^{+0.28}$ & 0.34$~\pm~$0.02 \\
        IRAS~05020--2941 &  $11.38_{-2.60}^{+0.32}$ &  $12.49_{-1.04}^{+0.13}$ & $6.08_{-0.24}^{+1.22}$ & 0.62$~\pm~$0.05 \\
       IRAS~F06076--2139 & $11.15_{-4.81}^{+-0.79}$ &   $2.14_{-0.45}^{+0.30}$ & $2.35_{-0.00}^{+0.18}$ & 0.19$~\pm~$0.04 \\
        IRAS~06206--6315 &  $0.02_{-0}^{+1.74}$ &   $2.36_{-0.29}^{+0.56}$ & $3.20_{-0.14}^{+0.17}$ & 0.70$~\pm~$0.04 \\
        IRAS~06301--7934 &  $13.41_{-3.03}^{+0.46}$ &   $7.25_{-0.56}^{+0.62}$ & $5.91_{-0.00}^{+3.05}$ & 0.37$~\pm~$0.07 \\
        IRAS~06361--6217 &   $0.83_{-0.68}^{+0.48}$ &   $5.00_{-0.11}^{+0.65}$ & $3.52_{-0.14}^{+0.07}$ & 0.10$~\pm~$0.02 \\
        IRAS~07251--0248 &   $0.17_{-0.12}^{+0.28}$ &   $1.73_{-0.32}^{+0.25}$ & $4.33_{-0.26}^{+0.07}$ & 0.93$~\pm~$0.03 \\
         IRAS~08201+2801 &   $0.61_{-0.20}^{+1.75}$ &   $7.30_{-0.68}^{+0.63}$ & $4.80_{-0.48}^{+0.68}$ & 0.59$~\pm~$0.07 \\
       IRAS~F08520--6850 &  $0.02_{-0}^{+0.50}$ &   $8.06_{-0.36}^{+0.58}$ & $4.97_{-0.03}^{+0.37}$ & 0.16$~\pm~$0.01 \\
         IRAS~08572+3915 &   $0.06_{-0.03}^{+0.19}$ &   $9.75_{-0.22}^{+0.27}$ & $4.13_{-0.02}^{+0.04}$ & 0.13$~\pm~$0.01 \\
         IRAS~09539+0857 &   $2.89_{-0.28}^{+2.03}$ &   $9.45_{-0.74}^{+0.54}$ & $6.22_{-0.32}^{+0.61}$ & 0.41$~\pm~$0.05 \\
       IRAS~F10038--3338 &   $6.06_{-0.65}^{+0.80}$ &  $11.20_{-0.16}^{+0.40}$ & $6.76_{-0.18}^{+0.22}$ & 0.40$~\pm~$0.01 \\
         IRAS~10091+4704 &   $2.94_{-1.04}^{+0.89}$ &   $3.89_{-0.95}^{+0.48}$ & $6.15_{-0.38}^{+0.20}$ & 0.62$~\pm~$0.08 \\
         IRAS~10173+0828 &  $19.29_{-2.38}^{+3.85}$ &   $5.39_{-0.52}^{+0.42}$ & $5.19_{-1.25}^{+0.00}$ & 0.26$~\pm~$0.16 \\
        IRAS~F10237+4720 &  $13.68_{-2.80}^{+0.04}$ &   $6.63_{-0.23}^{+2.15}$ & $7.18_{-0.00}^{+1.41}$ & 0.39$~\pm~$0.03 \\
         IRAS~10378+1109 &   $5.99_{-1.25}^{+1.72}$ &   $3.24_{-0.35}^{+0.70}$ & $3.44_{-0.08}^{+0.16}$ & 0.90$~\pm~$0.04 \\
        IRAS~10485--1447 &   $1.15_{-0.86}^{+0.73}$ &   $2.35_{-0.40}^{+0.23}$ & $3.73_{-0.30}^{+0.08}$ & 0.71$~\pm~$0.06 \\
         IRAS~11028+3130 &   $8.92_{-1.90}^{+4.56}$ &   $6.22_{-0.74}^{+1.13}$ & $5.97_{-0.31}^{+3.13}$ & 0.54$~\pm~$0.20 \\
         IRAS~11038+3217 &  $10.31_{-4.53}^{+2.40}$ &   $5.41_{-1.21}^{+1.01}$ & $4.89_{-0.00}^{+2.28}$ & 0.25$~\pm~$0.10 \\
        IRAS~11095--0238 &   $0.89_{-0.77}^{+0.10}$ &   $8.21_{-0.13}^{+0.57}$ & $4.96_{-0.00}^{+0.87}$ & 0.40$~\pm~$0.02 \\
        IRAS~11130--2659 &  $10.50_{-2.97}^{+1.16}$ &  $11.25_{-0.75}^{+0.46}$ & $6.06_{-0.00}^{+2.48}$ & 0.40$~\pm~$0.05 \\
         IRAS~11180+1623 &   $9.86_{-4.40}^{+3.07}$ &   $9.71_{-1.64}^{+0.41}$ & $6.04_{-0.64}^{+0.10}$ & 0.60$~\pm~$0.12 \\
        IRAS~11223--1244 &   $0.48_{-0.12}^{+2.99}$ &   $2.31_{-1.13}^{+0.62}$ & $3.13_{-0.22}^{+0.23}$ & 0.54$~\pm~$0.06 \\
         IRAS~11506+1331 &   $0.18_{-0.04}^{+1.36}$ &   $8.80_{-0.59}^{+0.47}$ & $2.92_{-0.13}^{+0.20}$ & 0.33$~\pm~$0.03 \\
         IRAS~11524+1058 &  $10.18_{-4.37}^{+5.69}$ &   $7.17_{-0.87}^{+1.19}$ & $5.73_{-0.00}^{+2.13}$ & 0.60$~\pm~$0.19 \\
         IRAS~11582+3020 &   $5.47_{-1.48}^{+0.45}$ &   $6.05_{-1.03}^{+0.34}$ & $5.28_{-0.22}^{+0.53}$ & 0.32$~\pm~$0.03 \\
         IRAS~12032+1707 &  $0.02_{-0}^{+0.19}$ &   $2.94_{-0.17}^{+0.97}$ & $4.42_{-0.14}^{+0.18}$ & 0.52$~\pm~$0.03 \\
        IRAS~12127--1412 &   $5.08_{-0.40}^{+1.70}$ &   $0.65_{-0.24}^{+0.32}$ & $2.43_{-0.07}^{+0.05}$ & 0.35$~\pm~$0.01 \\
       IRAS~F12224--0624 &   $0.47_{-0.33}^{+0.53}$ &  $2.58_{-0.84}^{+-0.22}$ & $6.81_{-0.00}^{+1.13}$ & 1.10$~\pm~$0.04 \\
                NGC~4418 &   $8.90_{-0.37}^{+0.08}$ &  $11.34_{-0.16}^{+0.10}$ & $4.53_{-0.02}^{+0.03}$ & 0.62$~\pm~$0.01 \\
        IRAS~12359--0725 &   $3.13_{-2.18}^{+2.73}$ &   $5.52_{-0.62}^{+0.93}$ & $3.08_{-0.24}^{+0.15}$ & 0.38$~\pm~$0.07 \\
         IRAS~12447+3721 &  $24.43_{-2.66}^{+4.72}$ &   $6.36_{-0.54}^{+0.99}$ & $4.48_{-0.36}^{+0.71}$ & 0.00$~\pm~$0.00 \\
        IRAS~F13045+2354 &  $0.02_{-0}^{+4.30}$ &   $5.66_{-1.23}^{+1.25}$ & $2.21_{-0.22}^{+0.08}$ & 0.17$~\pm~$0.05 \\
        IRAS~13106--0922 &   $0.05_{-0.01}^{+0.36}$ &   $1.97_{-0.70}^{+0.57}$ & $6.67_{-0.24}^{+0.12}$ & 1.20$~\pm~$0.05 \\
        IRAS~F13279+3401 &  $22.62_{-1.64}^{+1.43}$ &   $8.77_{-1.33}^{+0.38}$ & $6.39_{-0.00}^{+1.40}$ & 0.43$~\pm~$0.02 \\
         IRAS~13352+6402 &   $4.62_{-2.38}^{+1.05}$ &   $3.92_{-1.38}^{+0.11}$ & $3.70_{-0.05}^{+0.42}$ & 0.14$~\pm~$0.03 \\
                 Mrk~273 &  $12.88_{-0.99}^{+1.31}$ &   $5.04_{-0.49}^{+0.14}$ & $3.63_{-0.18}^{+0.08}$ & 0.53$~\pm~$0.02 \\
         IRAS~14070+0525 &   $0.31_{-0.22}^{+0.52}$ &   $2.91_{-0.81}^{+0.63}$ & $5.43_{-0.00}^{+1.18}$ & 0.73$~\pm~$0.05 \\
        IRAS~F14394+5332 &   $0.24_{-0.15}^{+0.64}$ &   $2.76_{-0.35}^{+0.58}$ & $3.01_{-0.26}^{+0.06}$ & 0.12$~\pm~$0.02 \\
        IRAS~F14511+1406 &   $0.47_{-0.33}^{+0.91}$ &  $0.02_{-0}^{+0.60}$ & $2.31_{-0.08}^{+0.12}$ & 0.21$~\pm~$0.02 \\
        IRAS~F14554+3858 &   $5.00_{-3.27}^{+4.50}$ &   $0.84_{-0.48}^{+1.89}$ & $3.71_{-0.14}^{+0.32}$ & 0.23$~\pm~$0.04 \\
         IRAS~15225+2350 &   $3.33_{-2.19}^{+1.20}$ &   $1.13_{-0.47}^{+0.36}$ & $3.06_{-0.00}^{+0.34}$ & 0.54$~\pm~$0.04 \\
         IRAS~15250+3609 &   $0.02_{-0.00}^{+0.15}$ &   $7.58_{-0.25}^{+0.22}$ & $4.72_{-0.09}^{+0.07}$ & 0.98$~\pm~$0.02 \\
                 Arp~220 &  $14.64_{-0.71}^{+0.61}$ &   $4.65_{-0.17}^{+0.13}$ & $4.82_{-0.23}^{+0.06}$ & 0.88$~\pm~$0.02 \\
FESS~J160655.82+541500.7 & $28.80_{-10.69}^{+3.70}$ &   $4.70_{-2.39}^{+3.89}$ & $1.61_{-0.00}^{+0.51}$ & 0.03$~\pm~$0.04 \\
        IRAS~F16073+0209 &   $0.06_{-0.00}^{+0.56}$ &  $10.95_{-0.00}^{+2.38}$ & $2.07_{-0.09}^{+0.18}$ & 0.00$~\pm~$0.00 \\
        IRAS~16090--0139 &   $0.12_{-0.09}^{+0.26}$ &   $2.60_{-0.39}^{+0.26}$ & $4.15_{-0.06}^{+0.20}$ & 0.57$~\pm~$0.02 \\
        IRAS~F16156+0146 &   $0.62_{-0.06}^{+2.65}$ &  $14.09_{-2.69}^{+0.12}$ & $3.46_{-0.06}^{+0.32}$ & 0.32$~\pm~$0.02 \\
        IRAS~F16242+2218 &  $0.22_{-0}^{+3.46}$ &  $1.91_{-0}^{+4.78}$ & $3.14_{-0.00}^{+1.03}$ & 0.53$~\pm~$0.21 \\
        IRAS~F16305+4823 &  $10.27_{-0.72}^{+1.64}$ &   $8.09_{-1.19}^{+0.50}$ & $5.68_{-0.27}^{+0.00}$ & 0.20$~\pm~$0.05 \\
         IRAS~16300+1558 &  $11.68_{-2.62}^{+0.38}$ &   $6.88_{-0.54}^{+0.31}$ & $5.49_{-0.00}^{+1.57}$ & 0.51$~\pm~$0.04 \\
         IRAS~16455+4553 &   $6.29_{-3.03}^{+4.08}$ &   $5.59_{-1.09}^{+0.11}$ & $2.75_{-0.47}^{+0.14}$ & 0.21$~\pm~$0.08 \\
        IRAS~16468+5200W &   $0.09_{-0.03}^{+0.58}$ &   $5.00_{-0.37}^{+0.20}$ & $3.98_{-0.10}^{+0.37}$ & 0.75$~\pm~$0.05 \\
        IRAS~16468+5200E &   $0.48_{-0.35}^{+0.67}$ &   $4.40_{-0.36}^{+0.31}$ & $4.19_{-0.23}^{+0.30}$ & 0.67$~\pm~$0.05 \\
         IRAS~17044+6720 &   $3.78_{-1.68}^{+0.42}$ &   $5.37_{-0.25}^{+0.51}$ & $2.00_{-0.07}^{+0.07}$ & 0.06$~\pm~$0.01 \\
        IRAS~F17028+3616 &  $39.11_{-3.83}^{+0.31}$ &   $9.57_{-1.56}^{+2.86}$ & $2.59_{-0.31}^{+0.07}$ & 0.37$~\pm~$0.09 \\
         IRAS~17068+4027 &   $2.61_{-0.24}^{+2.85}$ &   $9.16_{-0.58}^{+0.77}$ & $4.03_{-0.88}^{+0.00}$ & 0.51$~\pm~$0.03 \\
        IRAS~17208--0014 &  $17.17_{-2.67}^{+0.70}$ &   $5.94_{-0.25}^{+0.26}$ & $4.66_{-0.22}^{+0.12}$ & 0.52$~\pm~$0.03 \\
         IRAS~17463+5806 &  $16.88_{-1.02}^{+6.54}$ &   $8.96_{-0.30}^{+2.51}$ & $4.11_{-0.28}^{+0.03}$ & 0.38$~\pm~$0.07 \\
         IRAS~17540+2935 &  $10.50_{-3.61}^{+1.15}$ &   $2.94_{-0.60}^{+0.57}$ & $3.37_{-0.48}^{+0.00}$ & 0.23$~\pm~$0.06 \\
         IRAS~18443+7433 &   $5.20_{-1.56}^{+0.84}$ &   $6.76_{-0.62}^{+0.19}$ & $4.35_{-0.00}^{+0.63}$ & 0.33$~\pm~$0.03 \\
        IRAS~18531--4616 &   $2.80_{-0.82}^{+2.30}$ &   $5.67_{-1.54}^{+0.50}$ & $5.35_{-0.75}^{+0.00}$ & 0.20$~\pm~$0.07 \\
         IRAS~18588+3517 &  $15.90_{-3.11}^{+1.19}$ &   $4.90_{-0.25}^{+0.63}$ & $5.22_{-0.23}^{+0.71}$ & 0.47$~\pm~$0.05 \\
        IRAS~20100--4156 &   $0.71_{-0.58}^{+0.62}$ &   $2.94_{-0.28}^{+0.31}$ & $5.05_{-0.39}^{+0.02}$ & 1.24$~\pm~$0.04 \\
        IRAS~20109--3003 &  $15.07_{-3.02}^{+3.17}$ &   $8.19_{-0.90}^{+0.42}$ & $4.31_{-0.22}^{+0.29}$ & 0.76$~\pm~$0.29 \\
         IRAS~20286+1846 &   $0.53_{-0.17}^{+2.25}$ &   $2.07_{-0.84}^{+0.77}$ & $4.57_{-0.52}^{+0.44}$ & 0.73$~\pm~$0.13 \\
        IRAS~20551--4250 &   $0.91_{-0.77}^{+0.37}$ &   $6.99_{-0.64}^{+0.07}$ & $3.96_{-0.00}^{+0.68}$ & 0.26$~\pm~$0.02 \\
         IRAS~21077+3358 &  $0.19_{-0}^{+1.52}$ &   $2.21_{-0.22}^{+0.92}$ & $3.34_{-0.11}^{+0.29}$ & 0.11$~\pm~$0.06 \\
         IRAS~21272+2514 &   $4.43_{-2.70}^{+2.77}$ &   $2.97_{-0.76}^{+0.49}$ & $4.31_{-0.54}^{+0.32}$ & 0.88$~\pm~$0.10 \\
       IRAS~F21329--2346 &  $13.76_{-4.82}^{+2.28}$ &   $6.73_{-1.18}^{+0.17}$ & $5.42_{-0.87}^{+0.44}$ & 0.51$~\pm~$0.07 \\
       IRAS~22088--1831W &  $11.05_{-2.00}^{+1.09}$ &   $7.72_{-0.65}^{+0.54}$ & $5.07_{-0.74}^{+0.31}$ & 0.71$~\pm~$0.07 \\
       IRAS~22088--1831E &   $9.72_{-2.36}^{+0.49}$ &   $7.62_{-1.12}^{+0.17}$ & $5.08_{-0.00}^{+1.46}$ & 0.72$~\pm~$0.07 \\
         IRAS~22116+0437 &  $13.42_{-0.80}^{+3.79}$ &  $10.58_{-0.33}^{+1.13}$ & $4.78_{-0.45}^{+0.00}$ & 0.04$~\pm~$0.02 \\
                NGC~7479 &  $26.78_{-0.47}^{+1.32}$ &   $5.37_{-0.16}^{+0.52}$ & $2.30_{-0.10}^{+0.02}$ & 0.11$~\pm~$0.01 \\
         IRAS~23129+2548 &   $3.57_{-0.54}^{+0.51}$ &   $7.77_{-0.40}^{+0.35}$ & $5.24_{-0.11}^{+0.50}$ & 0.53$~\pm~$0.03 \\
        IRAS~F23234+0946 &   $5.72_{-1.39}^{+5.09}$ &   $1.85_{-0.44}^{+0.89}$ & $3.77_{-0.38}^{+0.10}$ & 0.33$~\pm~$0.07 \\
        IRAS~23230--6926 &   $8.07_{-1.52}^{+1.23}$ &   $7.29_{-0.46}^{+0.21}$ & $4.49_{-0.44}^{+0.32}$ & 0.72$~\pm~$0.04 \\
         IRAS~23365+3604 &   $4.98_{-2.57}^{+1.58}$ &   $0.35_{-0.02}^{+0.41}$ & $2.63_{-0.00}^{+0.31}$ & 0.42$~\pm~$0.09 \\
\enddata

\tablecomments{Column 1: the name of the object; Column 2: the mass fraction of amorphous pyroxene to total amorphous silicate; Column 3: the mass fraction of crystalline silicate to total silicate dust; Columns 4 and 5: the optical depths of the 10~$\mu$m silicate and the 6~$\mu$m $\rm H_2O$ ice absorption features, respectively, which are used in Sect.~\ref{sec:discuss}.}

\end{deluxetable*}

\bibliography{IRS_silicate}{}
\bibliographystyle{aasjournal}

\end{document}